\documentclass[aps,pra,reprint,10pt]{revtex4-2}

\usepackage{amsmath,amssymb}
\usepackage{mathtools}
\usepackage{bbold}
\usepackage{mathrsfs}

\usepackage{braket}
\usepackage{float}
\usepackage{cancel}

\usepackage{graphicx}
\usepackage{wrapfig}
\usepackage[caption=false]{subfig}
\usepackage{epstopdf}
\usepackage{multirow}

\usepackage{url}
\usepackage{comment}
\usepackage{lipsum}
\usepackage{overpic} 
\usepackage{xcolor}
\usepackage[framemethod=tikz]{mdframed}

\definecolor{lightgrey}{rgb}{0.8,0.8,0.8}
\definecolor{verylightgrey}{rgb}{0.9,0.9,0.9}

\newcommand{\Trace}{\mathrm{Tr}}
\newcommand{\Nsites}{\mathrm{N}}

\newcommand{\Ptrot}{\mathrm{P}}
\newcommand{\CD}{\scriptscriptstyle\mathrm{CD}}
\newcommand{\CDan}{\text{CD}^{\mathrm{an}}_{1}}
\newcommand{\CDvar}{\text{CD}^{\mathrm{var}}_{1}}

\newcommand{\opt}{\scriptscriptstyle\mathrm{opt}}

\newcommand{\res}{\mathrm{res}}
\newcommand{\targ}{\mathrm{targ}}
\newcommand{\drive}{\mathrm{drive}}

\newcommand{\ud}{\mathrm{d}}
\newcommand{\e}{\mathrm{e}}

\newcommand{\prm}{\mathrm{p}}
\newcommand{\Nfermions}{\mathrm{\hat{N}_F}}

\newcommand{\C}{{\mathbf C}}

\newcommand{\calA}{\mathcal{A}}
\newcommand{\calS}{\mathcal{S}}

\newcommand{\Nbasis}{\mathrm{N_c}}
\newcommand{\rmC}{{\mathrm C}}

\newcommand{\btheta}{\boldsymbol{\theta}}

\newcommand{\Ho}{\hat{H}}
\newcommand{\Uo}{\hat{U}}
\newcommand{\PauliSigma}{\hat{\sigma}}

\newcommand{\Hc}{\mathrm{H.c.}}

\newcommand{\opc}[1]{{\hat{c}}_{#1}}
\newcommand{\opcdag}[1]{{\hat{c}^{\dagger}}_{#1}}

\newcommand{\Deltat}{\Delta_{t}}
\newcommand{\Deltap}{\Delta_{p}}
\newcommand{\Pscript}{\scriptscriptstyle\mathrm{P}}

\usepackage{hyperref}

\begin{document}
\author{Kiran Thengil}
\email[Email (Corresponding Author):]{kthengil@sissa.it}
\affiliation{SISSA, Via Bonomea 265, I-34136 Trieste, Italy}

\author{Vincenzo Roberto Arezzo}
\affiliation{SISSA, Via Bonomea 265, I-34136 Trieste, Italy}

\author{Giuseppe E. Santoro}
\affiliation{SISSA, Via Bonomea 265, I-34136 Trieste, Italy}
\affiliation{International Centre for Theoretical Physics (ICTP), P.O.Box 586, I-34014 Trieste, Italy}

\title{Digital techniques for the frustrated Ising ring: the role of counter-diabatic terms}
\begin{abstract}
We investigate the role of local counter-diabatic (CD) terms in enhancing the performance of discrete-time digital protocols for a frustrated Ising ring, a system with an exponentially small spectral gap that acts as a bottleneck for conventional quantum annealing. 
The techniques investigated range from a digitised version of a fixed-schedule protocol, including lowest-order analytical CD terms, all the way to a full-fledged Quantum Approximate Optimisation Algorithm (QAOA), including variational CD terms (DC-QAOA).
By analyzing the resulting residual energy, we show that DC-QAOA combined with Chopped RAndom Basis (CRAB) quantum control techniques for optimizing the circuit parameters, can outperform all the other strategies. 
By monitoring the ground-state population during the dynamics 
we learn that DC-QAOA finds effective shortcuts towards the target state through intermediate excited states, a process that is unexpectedly enhanced by the inclusion of local CD unitaries. 
Our results highlight the importance of flexible variational control of CD dynamics and demonstrate that digital optimization can explore operator dynamics that remain inaccessible to standard analytical CD constructions.
\end{abstract}
\maketitle

\section{Introduction}
Quantum annealing~\cite{FinnilaQuantum1994,KadowakiQuantum1998,FarhiQuantum2000,GiuseppeTheory2002,GiuseppeOptimization2006,AlbashAdiabatic2018} (QA) is a paradigmatic framework for quantum optimization, in which the ground state of a desired target Hamiltonian $\Ho_{\targ}$ is constructed by slowly interpolating from an initial drive Hamiltonian $\Ho_{\drive}$ with a known and easily realizable ground state, $\Ho(s)=s \Ho_{\targ}+(1-s)\Ho_{\drive}$, with $s(t)$ --- the interpolation schedule --- starting from $s(0)=0$ and ending at $s(\tau)=1$ at the final time $\tau$. 
By encoding the solution of a computational or physical problem into $\Ho_\targ$, QA has been widely explored as an approach to tackle hard optimization problems and to study complex quantum many-body systems~\cite{DasColloquium2008,HaukeCan2012}. 
During a QA protocol, the presence of small energy gaps along the evolution requires long annealing times to maintain adiabaticity~\cite{CanevaAdiabatic2007,KnyshZero2016}.
In practice, however, annealing protocols must be executed within finite times due to decoherence, noise, and hardware constraints, placing them inherently outside the strictly adiabatic regime~\cite{PolkovnikovColloquium2011,ZancaQuantum2016}. 
As a result, non-adiabatic transitions are inevitably induced during the evolution, particularly near regions of small energy gaps or quantum critical points, leading to excitations and reduced target-state fidelity. Understanding and mitigating these finite-time, non-adiabatic effects is therefore central to improving the performance of QA, and has motivated the development of control strategies that go beyond the conventional adiabatic paradigm, collectively referred to as shortcuts to adiabaticity~\cite{TorronteguiShortcuts2013,OdelinShortcuts2019}. 
Within the shortcuts to adiabaticity framework, one of the most widely used strategies is counter-diabatic (CD) driving~\cite{DemirplakAssisted2005,BerryTransitionless2009}.

The fundamental idea of counter-diabaticity is to suppress the diabatic excitations which bring the system out of the instantaneous ground state, by incorporating additional interactions --- the so-called {\em adiabatic gauge potential} (AGP) $\hat{\calA}(s)$~\cite{KOLODRUBETZ20171} ---    
into a Hamiltonian $\Ho_{\CD}(s,\dot{s}) =\Ho(s)+\dot{s}\, \hat{\calA}(s)$. 
In principle, by calculating the complete $\hat{\calA}(s)$, the evolution is entirely confined to the adiabatic states of $\Ho(s)$. 
However, for many-body systems, achieving the complete $\hat{\calA}(s)$ requires detailed spectral information about the instantaneous Hamiltonian and typically yields highly non-local interactions that are difficult to implement. 
Initial efforts addressed this limitation in the continuous-time framework using local AGPs, obtained by analytically minimizing an appropriate action~\cite{SelsMinimizing2017}. 
Subsequently, Claeys \emph{et al.}~\cite{ClaeysFloquet2019} proposed a nested-commutator construction for the AGP with progressively increasing non-locality. This analytical optimization strategy was extended to discrete-time dynamics in Ref.~\cite{HegadeDigitized2022}, making it suitable for implementation on gate-based quantum devices. 
Alternatively, a numerical optimization strategy for AGP was proposed in Ref.~\cite{ChandaranaDigitized2022} --- digitized counter-diabatic quantum approximate optimization algorithm (DC-QAOA) --- and was shown to reduce overall circuit complexity. 
Over the years, these AGP-based methods have found applications across diverse physical systems—ranging from non-integrable spin-$1/2$ chains~\cite{ZhouExperimental2020} to NP-hard optimization problems such as bin packing~\cite{XuDigitized2025} and protein folding~\cite{RomeroBias2025}.

In this paper, we reconsider the frustrated Ising ring model proposed in Ref.~\cite{RobertsNoise2020} and discussed in the context of QA in Ref.~\cite{CoteDiabatic2023}, and with digitized techniques in Refs.~\cite{RuiyiExponential2025,ArezzoDigital2025}.
Conventional QA faces a major challenge in this system, as it exhibits a spin-glass-like bottleneck due to an exponentially small energy gap along the annealing path. 
For this system, a recent study~\cite{GrabaritsFighting2025} within a continuous-time framework reported that, for a fixed schedule $s(t)$, local CD terms help enlarge the minimum energy gap along the annealing path. 
However, even with an analytically optimised second-order AGP, the resulting protocol remains insufficient to achieve high-fidelity solutions.
Furthermore, it has been shown~\cite{Arezzo_arxiv2026_cont_time} that, within continuous-time dynamics, the inclusion of lowest-order counter-diabaticity terms provides no improvement if the schedule $s(t)$ is optimized.

Here, we conduct a systematic comparison of different discrete-time techniques for the frustrated Ising ring. Our focus is on understanding the influence of local CD terms on the discrete quantum dynamics and on the performance of the algorithm. 
To this end, we analyze several techniques with and without counter-diabatic terms, including: 0) the bare use of a smooth fixed schedule, 1) a fixed-schedule with analytically optimized lowest-order CD terms, 2) the use of an optimized smooth schedule, 3) an optimized schedule with analytically optimized lowest-order CD terms, 4) plain QAOA without CD, and 5) QAOA with variationally optimized CD terms (a.k.a. DC-QAOA\cite{ChandaranaDigitized2022}).
All these techniques are consistently interpreted within a single framework of repeatedly applied Trotterized unitary operations. 
For the numerical optimization of variational parameters in methods 2)--5), we use a Fourier-based quantum control technique, the Chopped RAndom Basis (CRAB)~\cite{RachDressing2015}, to obtain optimized smooth schedules with improved performance. 
For a fair comparison, we restrict the choice of the counter-diabatic terms to an AGP constructed with lowest-order nested-commutator.

The performance of each method is quantified by the closeness of the obtained solution to the exact ground state of the frustrated Ising ring.  
We observe that the limitations of analytically optimized counter-diabatic terms, previously reported in continuous-time dynamics~\cite{Arezzo_arxiv2026_cont_time}, remain in the corresponding discrete-time techniques. 
We show that this limitation 
can be overcome through a numerical optimization based on CRAB. 
Finally, a detailed analysis of the evolution path in terms of the instantaneous fidelity reveals the underlying mechanism behind the superiority of DC-QAOA: it realizes an efficient shortcut-to-adiabaticity by inducing excitations to higher energy levels at the very beginning of the evolution.

The paper is organized as follows. 
In Sec.~\ref{sec:methods} we discuss the system, including its Jordan--Wigner transformation, and the various digitized dynamics explored and compared in our work. 
The results obtained are presented and analyzed in Sec.~\ref{sec:Results}. 
Finally, we conclude the paper, in Sec.~\ref{sec:conclusions}, with a summary, a discussion and perspectives for future work. 
A few final appendices are included to clarify some technical aspects of our work. 

\section{System, dynamics, and methods}
\label{sec:methods}
\begin{figure}[H]
    \centering
    \includegraphics[width=0.9\columnwidth]{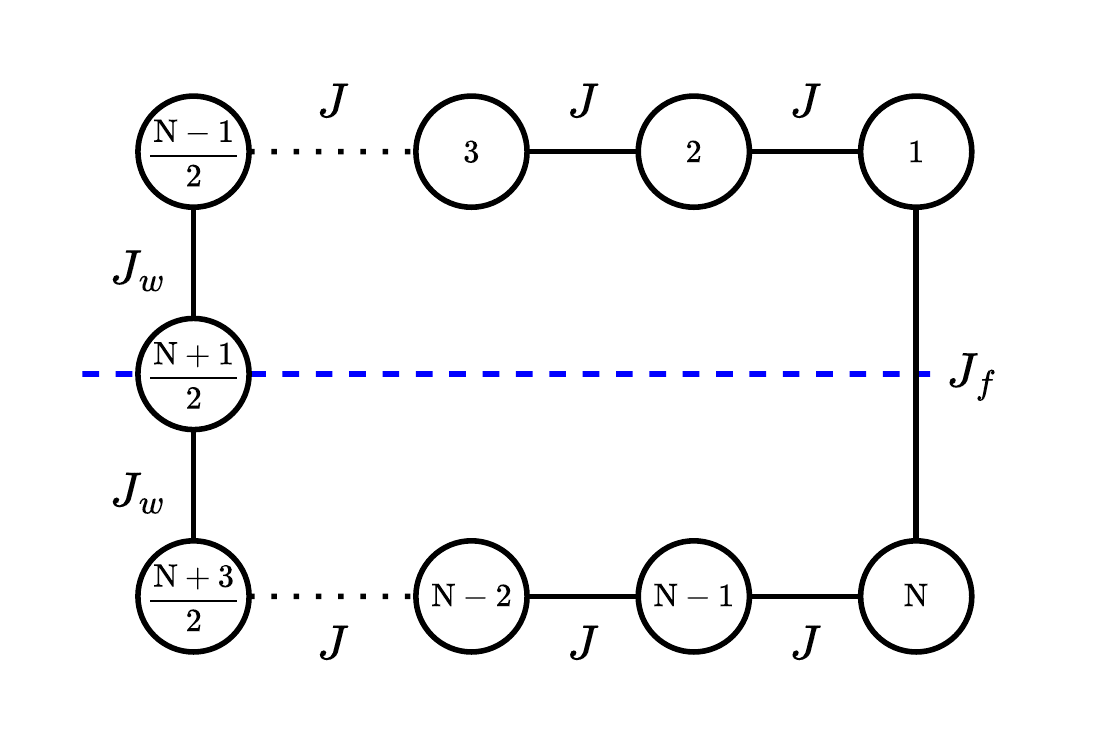}
    \caption{Schematic diagram of the frustrated Ising ring model with an odd number of sites $\Nsites$, ferromagnetic coupling $J$, weak ferromagnetic coupling $J_w$, and antiferromagnetic frustrating coupling $J_f$. This model is reflection symmetric about the blue dashed line.}
    \label{system}
\end{figure}

We consider a frustrated Ising ring model with an odd number of sites $\Nsites$. Here, the target Hamiltonian, whose ground state we aim at constructing, is:
\begin{equation}
    \Ho_{\targ} = \Ho_{z} = -\sum_{j=1}^{\Nsites}J_{j}\PauliSigma^{z}_{j}\PauliSigma^{z}_{j+1} \;,
    \label{eqn:Hz}
\end{equation}
where $\PauliSigma^{z}_j$ is the Pauli-Z operator at site $j$, and the nearest-neighbor couplings are
\begin{equation}
  J_{j} =
    \begin{cases}
      J_w & \text{if $j=(\Nsites\pm 1)/2$}\\
      -J_f & \text{if $j=\Nsites$}\\
      J & \text{otherwise}
    \end{cases}       \;.
\end{equation}
Most of the sites are coupled via ferromagnetic coupling $J$, except for two central ones, which have weaker ferromagnetic coupling $J_{w}$, and a single antiferromagnetic coupling $J_{f}$ that causes frustration in the ring; see Fig.~\ref{system}. 
In this paper, we choose $J=1$ as the unit of energy, $J_{w}=0.5$, and different values for $J_{f}$, such that $JJ_{f}>J^2_{w}$ leading to the spin-glass bottleneck discussed in recent studies~\cite{GrabaritsFighting2025,RobertsNoise2020,CoteDiabatic2023,RuiyiExponential2025}.

The driving Hamiltonian is the standard transverse field term
\begin{equation}
    \Ho_\drive = \Ho_x = -h \sum_{j=1}^\Nsites\PauliSigma^x_j \;,
\end{equation}
where $\PauliSigma^{x}$ is Pauli-X operator and $h$ is the uniform transverse field coupling. Using the Jordan-Wigner transformation \cite{MbengQuantum2019} one can map $\Ho_{x/z}$ into quadratic fermionic Hamiltonians:
\begin{equation} \label{eqn:Hzz_fermions}
\begin{aligned}
\Ho_z &= -\sum_{j=1}^{\Nsites} J_j 
\Big( \opcdag{j} \opc{j+1} \,+\, \opcdag{j} \opcdag{j+1}  \,+\, {\Hc} \Big),\\
\Ho_x &=  h\sum_{j=1}^{\Nsites} \Big( \opcdag{j} \opc{j} - \opc{j} \opcdag{j} \Big) \;,
\end{aligned}
\end{equation}
where $\hat{c}^\dagger_j$ and $\hat{c}_j$ are fermionic creation and annihilation operators, 
with the boundary condition $\opc{\Nsites+1}=(-1)^{\prm+1}\opc{1}$.
Here, $\prm=0$ or $1$ denotes the even or odd fermion-parity sectors of the Hilbert space,
with $(-1)^{\prm}=\e^{i\pi \Nfermions}$, where
$\Nfermions=\sum_j \opcdag{j}\opc{j}$ is the total number of fermions.

The ground state of $\Ho_{x}$ for $h<0$ is the fully occupied fermionic state
$|\psi_0\rangle=\prod_j \opcdag{j}|0\rangle$, which has $\prm=1$ when $\Nsites$ is odd.
We fix $h=-J$ and restrict ourselves to the odd fermion-parity sector, where
$\hat{c}_{\Nsites+1}=\hat{c}_1$. 
Since both $\Ho_x$ and $\Ho_z$ present reflection symmetry ($J_j = J_{\Nsites-j}$), as shown in Fig.\ref{system}, it is then convenient to perform the transformation introduced in Ref.~\cite{GrabaritsFighting2025}.
Further details on this transformation, which reduces the computational complexity from dealing with $2\Nsites\times 2\Nsites$ matrices to more convenient $\Nsites\times \Nsites$ ones, can be found in Appendix A of Ref.~\cite{Arezzo_arxiv2026_cont_time}. 

\begin{figure}[ht]
    \centering
    \subfloat[]{\includegraphics[width=0.9\columnwidth]{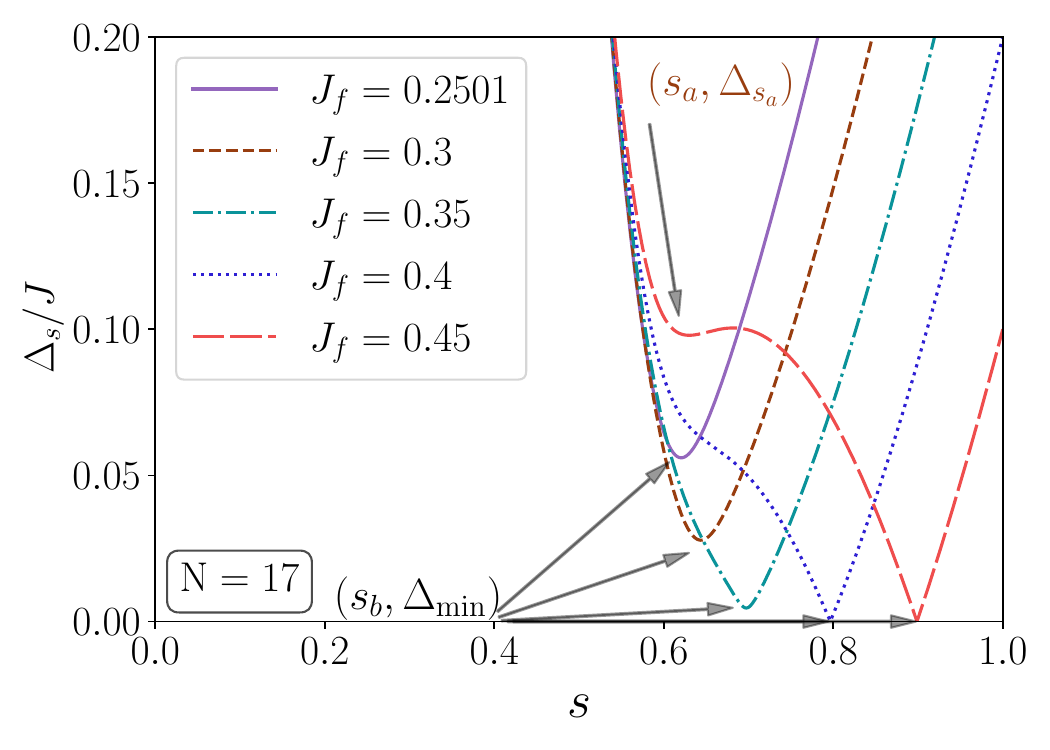}\label{EnergyGapVsSchedule_N_17}}
    \hfill
    \subfloat[]{\includegraphics[width=0.9\columnwidth]{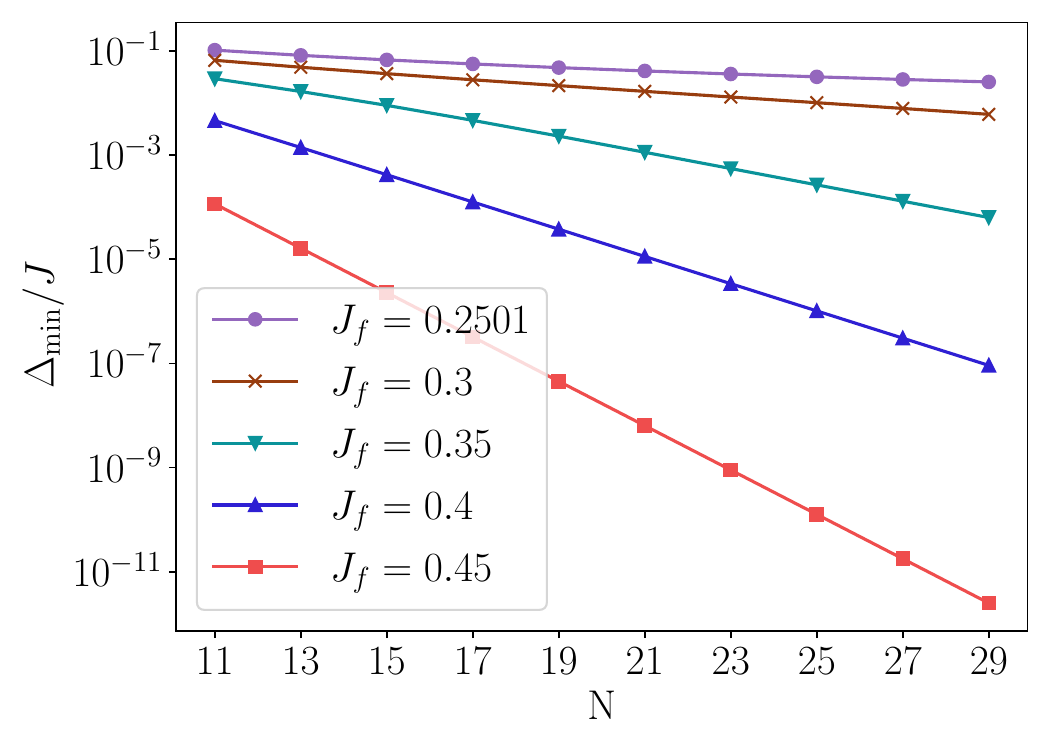}\label{MinEneVsN}}
    \caption{(a) 
    The energy gap $\Delta_{s}/J$ between the ground state and the first excited state of the interpolating Hamiltonian $\Ho(s)$ in Eq.~\eqref{InterpHamil} at different values of $s\in [0,1]$ and for various frustrating couplings $J_f$. Here, the system size is $\Nsites=17$. The minimum value of the energy gaps and its corresponding $s$ are indicated as $(s_b,\Delta_{\mathrm{min}})$. Also, we have explicitly marked the second small gap for $J_{f}=0.45$ using $(s_a,\Delta_{s_a})$. 
    (b) Exponential decrease of the minimum energy gap $\Delta_{\mathrm{min}}/J$ between the ground state and the first excited state of $\Ho(s)$ as the number of spins $\Nsites$ increases, for different values of the antiferromagnetic coupling $J_f$.}
    \label{MinE_gapVsJf}
\end{figure}

A continuous-time QA dynamics requires solving the Schr\"odinger equation
\begin{equation}
    i\hbar \frac{\ud}{\ud t} |\psi(t)\rangle = \Ho(s(t)) |\psi(t)\rangle \;, 
\end{equation}
where the interpolating Hamiltonian $\Ho(s)$ reads:
\begin{equation}
    \Ho(s) = s \Ho_{z}+(1-s)\Ho_{x} \;.
    \label{InterpHamil}
\end{equation}
Here the schedule $s(t)$ is an arbitrary function of time with boundary conditions $s(0) = 0$ and $s(\tau) = 1$, $\tau$ being the total annealing time.
The instantaneous energy gap $\Delta_s$ between the ground state and first excited state of the interpolating Hamiltonian $\Ho(s)$ for values of $s\in[0,1]$ is shown in Fig.~\ref{EnergyGapVsSchedule_N_17}, for a chain of $\Nsites=17$ sites with different values of the frustration $J_f$. 
The minimum energy gap and the corresponding $s$ values are indicated using the ordered pair $(s_b,\Delta_{\mathrm{min}})$. The weakly frustrated systems posses a single minimum energy gap, while strong frustration leads to an additional small gap other than the minimum gap. This additional small gap for $J_f=0.45$ is particularly indicated using $\Delta_{s_a}\approx 0.098~J$ at $s_a\approx 0.63$. 
Furthermore, the minimum energy gap shifts towards higher values of $s$ and approaches smaller and smaller values with increasing frustration. 
Figure \ref{MinEneVsN} shows the exponential decrease of the $\Delta_{\mathrm{min}}$ with increasing system size for different values $J_f$. The linear behavior of $\Delta_{\mathrm{min}}$ in logarithmic scale confirms its exponential decrease with increasing $\Nsites$ causing the spin-glass bottleneck. The minimum time duration required for the adiabatic dynamics of the interpolating Hamiltonian can be approximated as~\cite{GrabaritsFighting2025}
\begin{equation}   \tau_{\mathrm{ad}}\approx\Delta_{\mathrm{min}}^{-2} \;,
\end{equation}
which increases exponentially with $\Nsites$ for the system under consideration. Also, increasing the frustrating coupling strength $J_f$ leads to a much steeper decrease of the minimum energy gap, creating a significant challenge to conventional QA, even in moderately sized systems.

Based on the quantum adiabatic theorem~\cite{Morita_JMP2008}, 
for $\tau>>\tau_{\mathrm{ad}}$, QA attains the ground state of $\Ho_{z}$ for a system initialized in the ground state $\ket{\psi_{0}}$ of $\Ho_{x}$. 
This is true for an arbitrary $s(t)$ in the adiabatic limit $\tau\to \infty$.
However, in finite-$\tau$ QA dynamics, the final evolved state $|\psi(\tau)\rangle$ depends strongly on the functional form of $s(t)$ and generally may not approach the required target state. 

Definitely, the simplest schedule one can think of is the linear one:
\begin{equation} \label{linearrampeqn}
    s^{(\mathrm{lin})}(t) = \frac{t}{\tau} \;.
\end{equation} 
Based on arguments relying on the adiabatic theorem~\cite{Morita_JMP2008}, it has been suggested that schedules with a vanishing derivative at $t=0$ and $t=\tau$ should yield better asymptotic performance. Among those, two popular choices have been the 
cubic ramp~\cite{GrabaritsFighting2025} 
\begin{equation} \label{cubicrampeqn}
    s^{(\mathrm{cub})}(t) = 3\left(\frac{t}{\tau} \right)^2-2\left(\frac{t}{\tau} \right)^3 \;,
\end{equation} 
and a sinusoidal one~\cite{HegadeDigitized2022,IevaCounterdiabatic2023}
\begin{equation} \label{sinerampeqn}
    s^{(\mathrm{sin})}(t) = \sin^{2}\left[\frac{\pi}{2}\sin^{2}\left(\frac{\pi t}{2 \tau}\right)\right] \;.
\end{equation}
An alternative to those fixed schedules would be to ``decorate'' them with some extra variational freedom. For instance, starting from the linear schedule, one might add arbitrary Fourier components to $s(t)$ by writing:
\begin{equation} \label{eqn:s_Fourier}
    s(t)  = \frac{t}{\tau}\Big(1 + \sum_{n} \rmC_n \sin(\omega_n (t-\tau)) \Big) \;,
\end{equation}
with appropriate frequencies $\omega_n$, in principle, an infinite number of them: this is the basis of a quantum control technique known as Chopped RAndom Basis (CRAB)\cite{Caneva_PRA2011}, where a {\em finite} number of frequencies $\omega_n$ are introduced, and the corresponding coefficients $C_n$ optimized. 

On the technical side, it is convenient to move from the continuous-time framework to a Trotter-digitized evolution. By dividing the total evolution time $\tau$ into $\Ptrot$ time-steps $\Deltat=\tau/\Ptrot$, one could write, to lowest-order in the Trotter splitting, a digitized QA (dQA) dynamics based on the evolution operator~\cite{MbengOptimal2019}:
\begin{equation} \label{eqn:U_dQA}
\Uo_{\Pscript}^{\scriptscriptstyle{\mathrm{dQA}}} = \prod_{p=1}^{\Ptrot}
    \e^{-i(1-s_p)\frac{\Deltat}{\hbar}\Ho_{x}}
    \e^{-is_p\frac{\Deltat}{\hbar}\Ho_{z}} \;,
\end{equation}
where $s_{p}=s(t_p)$ with $t_p=(p-\frac{1}{2})\Deltat$, in the middle of each $\Deltat$-interval. 
This, in turn, is a particular case of a more general variational {\em Ansatz}, known as Quantum Approximate Optimization Algorithm (QAOA)~\cite{FarhiA2014}:
\begin{equation} \label{eqn:U_QAOA}
\Uo_{\Pscript}^{\scriptscriptstyle{\mathrm{QAOA}}} = \prod_{p=1}^{\Ptrot}
    \e^{-i\theta^x_p \Ho_{x}}
    \e^{-i\theta^z_p \Ho_{z}} \;,
\end{equation}
when $\theta^x_p = (1-s_p) \frac{\Deltat}{\hbar}$ and $\theta^z_p = s_p \frac{\Deltat}{\hbar}$, leading to a final state given by:
\begin{equation} \label{eqn:psi_P_QAOA}
|\psi_{\Ptrot}(\btheta)\rangle = \Uo_{\Pscript}^{\scriptscriptstyle{\mathrm{QAOA}}}(\btheta) |\psi_0 \rangle \;.
\end{equation}
Here $\btheta = \left( \theta^{x}_{1},\theta^{z}_{1},\cdots,\theta^{x}_{\Pscript},\theta^{z}_{\Pscript}\right)$ denote the $2\Ptrot$ parameters appearing in the general QAOA {\em Ansatz}.

For the frustrated Ising ring model we are considering, when translated in terms of Jordan-Wigner fermions, the dynamics occurs within the subspace of fermionic Gaussian states, which can be expressed using the \emph{Thouless formula} \cite{Thouless_NP1960}:
\begin{equation}
\label{eqn:gaussianstate}
    \ket{\psi_{\mathrm{Gaussian}}} = \mathcal{N}\exp\left(\frac{1}{2}\sum_{jj'}^{\Nsites}\mathbf{Z}_{jj'}\opcdag{j}\opcdag{j'}\right) \ket{0},
\end{equation}
where $\mathcal{N}$ is a normalization factor, $\mathbf{Z}_{jj'}$ is a $\Nsites \times \Nsites$ complex antisymmetric matrix \cite{Thouless_NP1960,mbeng2024quantum} that characterizes the state, and $\ket{0}$ is the vacuum of the $\opcdag{j}$ operators. 
In the presence of a further lattice reflection symmetry, one can prove~\cite{ArezzoDigital2025} that the dimensionality of such a subspace is 
\begin{equation} \label{eqn:Pcr_symm}
\dim_{\mathrm{Gaussian}}^{\text{symm}}  =
\frac{\Nsites^2 -1}{2}  \hspace{5mm} \text{if} \ \Nsites \; \text{is odd} \;. 
\end{equation}
This suggests that our system should be fully {\em controllable}, i.e., the target state exactly reached, provided the number $N_{\btheta}$ of parameters $\btheta$ used in the parametrization of the evolution operator 
is at least as large as $\dim_{\mathrm{Gaussian}}^{\text{symm}}$, 
in the present case:
\begin{equation} \label{eqn:control_2P}
N_{\btheta} = 2\Ptrot \ge N_{\btheta}^{\mathrm{cr}} =
\frac{\Nsites^2 -1}{2} \;.
\end{equation}
Obviously, it is still interesting how to devise good algorithms for $\Ptrot$ well below this critical threshold of full controllability.

In that respect, it is worth considering an extension of this standard QA/QAOA framework which proposes using {\em counter-diabatic} techniques~\cite{HegadeDigitized2022}.

The exact counter-diabatic Hamiltonian for the interpolating Hamiltonian in Eq.~\eqref{InterpHamil} is obtained by the addition of an auxiliary adiabatic gauge potential:
\begin{equation}
    \Ho_{\CD}(t)=\Ho(s(t)) + \dot{s}(t) \, \hat{\calA}_{s(t)} \;,
\end{equation}
where the auxiliary part $\calA_{s(t)}$ is meant to reduce the diabatic transitions caused by $\Ho(s)$.
For many-body systems, an exact CD Hamiltonian consists of highly non-local interactions, which are hard to generate. 
Also, this requires, in general, complete spectral information on the interpolating Hamiltonian. 
As a countermeasure, recent studies considered  
approximate adiabatic gauge potentials (AGP)~\cite{ClaeysFloquet2019} with terms of increasing non-local interactions~\cite{ChandaranaDigitized2022}:
\begin{equation}
    \calA^{(\ell)}_{s}(t) = i \sum_{n=1}^{\ell} \alpha_{n}
    \underbrace{\bigl[ \Ho(s), \cdots , \bigl[ \Ho(s)}_{2n-1~\text{times}}, \tfrac{\partial}{\partial s}\Ho(s) \bigr]\bigr] \,.
    \label{AGP}
\end{equation}
Here, $\ell$ denotes the highest order retained in the truncated expansion of $\calA^{(\ell)}_{s}(t)$; the expansion consists of terms involving an odd number ($2n-1$) of nested commutators~\cite{ClaeysFloquet2019}, each weighted by a coefficient $\alpha_{n}$. To circumvent the complications arising from highly nonlocal interactions, we limit our consideration to the first-order AGP, taking:
\begin{equation}
\calA^{(1)}_{s} = i\alpha_1\bigl[ \Ho(s), \tfrac{\partial}{\partial s}\Ho(s) \bigr]
        \equiv \alpha_1\Ho_{xz} \;,
\label{AGP1st}
\end{equation} 
where 
\begin{equation}
\begin{aligned}
 \Ho_{xz} &=  i [\Ho_x , \Ho_z] \\
&= - 2  \sum_{j=1}^N J_j \Big( \PauliSigma^x_j \PauliSigma^y_{j+1} + \PauliSigma^y_j \PauliSigma^x_{j+1} \Big)  \nonumber \\ 
&=   - 4i  \sum_{j=1}^N J_j \Big( \opcdag{j} \opcdag{j+1} - \opc{j+1} \opc{j} \Big) 
\;,
\end{aligned}
\end{equation}
the final expression being in terms of Jordan-Wigner fermions. 
The inclusion of such lowest-order AGP term suggests generalizing the QAOA {\em Ansatz} by adding a further unitary to the Trotterized evolution operator~\cite{HegadeDigitized2022}:
\begin{equation} \label{eqn:U_CD}
\Uo^{\CD}_{\Pscript}(\btheta) = \prod_{p=1}^{\Ptrot}\e^{-i\theta^{xz}_{p}\Ho_{xz}}
\e^{-i\theta^x_{p}\Ho_x}
\e^{-i\theta^z_{p}\Ho_z} \;, 
\end{equation}
with the final state {\em Ansatz} involving now $3\Ptrot$ parameters 
$\btheta = \left( \theta^{x}_{1},\theta^{z}_{1},\theta^{xz}_{1},\cdots,\theta^{x}_{\Pscript},\theta^{z}_{\Pscript},\theta^{xz}_{\Pscript}\right)$:
\begin{equation} \label{eqn:psi_P_CD}
|\psi_{\Ptrot}(\btheta)\rangle = \Uo_{\Pscript}^{\CD}(\btheta) |\psi_0 \rangle \;.
\end{equation}
The controllability criterion in Eq.~\eqref{eqn:control_2P} would now read:
\begin{equation} \label{eqn:control_3P}
N_{\btheta} = 3\Ptrot \ge N_{\btheta}^{\mathrm{cr}} =
\frac{\Nsites^2 -1}{2} \;.
\end{equation}

Different optimization techniques can be realized depending on whether these
variational parameters are fixed \emph{a priori} or optimized. 
We refer the reader to App.~\ref{app:details} for further details on the various methods, limiting ourselves here to the main ingredients. 
In particular, we have studied and compared six different methods:
\begin{description}
\item[0) Fixed schedule] Definitely the simplest ``0th-order'' approach consists in using a fixed schedule $s(t)$, for instance $s^{\mathrm{lin}}(t)$ as in standard QA approaches, or improved versions with vanishing derivatives at $t=0$ and $t=\tau$, like $s^{\mathrm{cub}}(t)$ or $s^{\mathrm{sin}}(t)$, to deduce $s_p=s(t_p)$ at the discrete times $t_p=(p-\frac{1}{2})\Deltat$ with a fixed small $\Deltat$, and set in Eqs.~\eqref{eqn:U_CD}-\eqref{eqn:psi_P_CD}:
\begin{equation}
\left\{
\begin{array}{rcl}
s_p &=& s(t)|_{t=t_p} \vspace{3mm} \\
\theta^{z}_{p} &=& s_p\Deltat \vspace{3mm} \\
\theta^{x}_{p} &=& (1-s_p)\Deltat\, \vspace{3mm}\\
\theta^{xz}_{p} &=& 0
\end{array}
\right. \,.
\end{equation}
\item[1) Fixed schedule + analytical CD] A second possibility is to use a fixed schedule with a fixed small $\Deltat$, for instance $s^{\mathrm{cub}}(t)$, and include the CD term by analytically optimizing the coefficient $\alpha_1$ via the techniques explained in Ref.~\cite{KOLODRUBETZ20171}, which lead to
\begin{align}
\alpha^{\opt}_{1}(s_p) 
&= \frac{\hbar \, \Trace ([\Ho(s_p),\partial_s\Ho])^2}{\Trace([\Ho(s_p),[\Ho(s_p),\partial_s\Ho]])^2} \;.
\end{align}
This leads setting in Eqs.~\eqref{eqn:U_CD}-\eqref{eqn:psi_P_CD}:
\begin{equation}
\left\{
\begin{array}{rcl}
s_p &=& s^{\mathrm{cub}}(t)|_{t=t_p} \vspace{3mm} \\
\dot{s}_p &=& \dot{s}^{\mathrm{cub}}(t)|_{t=t_p} \vspace{3mm} \\
\theta^{z}_{p} &=& s_p\Deltat \vspace{3mm} \\
\theta^{x}_{p} &=& (1-s_p)\Deltat\, \vspace{3mm}\\
\theta^{xz}_{p} &=& \dot{s}_{p}\alpha^{\opt}_{1}(s_{p})\Deltat
\end{array}
\right. \,.
\end{equation}
Such an approach, with a further addition of longitudinal bias fields, is known in the literature as Digitized counter-diabatic Quantum Optimization (DCQO)~\cite{HegadeDigitized2022}, a method applied to a wide range of systems, including Ising spin glass models~\cite{HegadeDigitized2022}, HUBO problems~\cite{RomeroBias2025}, and Portfolio optimization~\cite{CadavidEfficient2024}.
In our study, we will not resort to adding longitudinal bias fields for two main reasons: first, because the ground state is a simple ferromagnet, and second, because it would ruin the Jordan-Wigner mapping to fermions which we will use. 
\item[2) Optimized schedule without CD]
Alternatively, one might consider a CRAB-based variational optimization of the schedule $s_p$ without any CD term, leading to:
\begin{equation}
\label{eqn:theta_Opt_sp}
\left\{
\begin{array}{rcl}
s_p &=& s^{\scriptscriptstyle\mathrm{CRAB}}(t)|_{t=t_p} \vspace{3mm} \\
\theta^{z}_{p} &=& s_p\Deltat \vspace{3mm} \\
\theta^{x}_{p} &=& (1-s_p)\Deltat\, \vspace{3mm}\\
\theta^{xz}_{p} &=& 0 
\end{array}
\right. \,.
\end{equation}
Here the schedule $s^{\scriptscriptstyle\mathrm{CRAB}}(t)$ is obtained by adding a finite number $\Nbasis$ of Fourier modes to the linear ramp $s^{(\mathrm{lin})}(t)=t/\tau$:
\begin{equation} \label{eqn:s_p_CRAB}
s^{\scriptscriptstyle\mathrm{CRAB}}(t)
= \frac{t}{\tau}
\Big(1 +\sum_{n=1}^{\Nbasis} \rmC_n \sin(\omega_n (t-\tau))\Big) \;,
\end{equation}
which has $\Nbasis$ free parameters $\C=(\rmC_1,....,\rmC_\Nbasis)$. 
\item[3) Optimized schedule + analytical CD]
As a further possible improvement, one might consider adding an analytical CD term, thereby writing
\begin{equation}
\left\{
\begin{array}{rcl}
s_p &=& s^{\scriptscriptstyle\mathrm{CRAB}}(t)|_{t=t_p} \vspace{3mm} \\
\dot{s}_p &=& \dot{s}^{\scriptscriptstyle\mathrm{CRAB}}(t)|_{t=t_p} \vspace{3mm} \\
\theta^{z}_{p} &=& s_p\Deltat \vspace{3mm} \\
\theta^{x}_{p} &=& (1-s_p)\Deltat\, \vspace{3mm}\\
\theta^{xz}_{p} &=& \dot{s}_{p}\alpha^{\opt}_{1}(s_{p})\Deltat 
\end{array}
\right. \,,
\end{equation}
with $s_p$ given by Eq.~\eqref{eqn:s_p_CRAB} and $\dot{s}_p$ calculated using the same CRAB schedule $s^{\scriptscriptstyle\mathrm{CRAB}}(t)$.
\item[4) QAOA without CD terms] 
A further alternative approach without CD is a full QAOA with angles smoothly optimized with CRAB~\cite{RuiyiExponential2025}.
Here we write:
\begin{equation}
\label{eqn:theta_QAOA}
\left\{
\begin{array}{rcl}    
\theta^{z}_p &=& \frac{t_p}{\tau}\Big(\rmC^{z}_0
+ \displaystyle \sum_{n=1}^{\Nbasis} \rmC^{z}_n 
    \sin(\omega_n (t_p-\tau)) \Big)
    \vspace{4mm} \\
\theta^{x}_p &=& \Big(1-\frac{t_p}{\tau}\Big) \Big(\rmC^{x}_0
+ \displaystyle{\sum_{n=1}^{\Nbasis}} 
\rmC^{x}_n \sin(\omega_n (t_p-\tau))\Big) 
\vspace{4mm} \\
\theta^{xz}_p &=& 0 
\end{array}
\right. ,
\end{equation}
with possible further dressed-CRAB refinements~\cite{RachDressing2015} detailed in App.~\ref{app:details}. This {\em Ansatz} involves $2\Nbasis+2$ free parameters  $\C=(\rmC_{0}^x,...,\rmC_{\Nbasis}^x,\rmC_{0}^z,...,\rmC_{\Nbasis}^z)$.
\item[5) QAOA with CD terms]
Finally, we can further add a variational CD term without relying on the analytic expression $\theta^{xz}_p=\dot{s}_{p}\alpha^{\opt}_{1}(s_{p})\Deltat$ with a choice
\begin{equation}
\label{eqn:theta_QAOA_CD}
\left\{
\begin{array}{rcl}    
\theta^{z}_p &=& \frac{t_p}{\tau}\Big(\rmC^{z}_0
+ \displaystyle \sum_{n=1}^{\Nbasis} \rmC^{z}_n 
    \sin(\omega_n (t_p-\tau)) \Big)
    \vspace{4mm} \\
\theta^{x}_p &=& \Big(1-\frac{t_p}{\tau}\Big) \Big(\rmC^{x}_0
+ \displaystyle{\sum_{n=1}^{\Nbasis}} 
\rmC^{x}_n \sin(\omega_n (t_p-\tau))\Big) 
\vspace{4mm} \\
\theta^{xz}_p &=& \rmC^{xz}_0 + 
\displaystyle \sum_{n=1}^{\Nbasis} \rmC^{xz}_n \sin(\omega_n (t_p-\tau))
\end{array}
\right. .
\end{equation}
This {\em Ansatz} involves $3\Nbasis+3$ free parameters $\C=(\rmC_{0}^x,...,\rmC_{\Nbasis}^x,\rmC_{0}^z,...,\rmC_{\Nbasis}^z,\rmC_{0}^{xz},...,\rmC_{\Nbasis}^{xz})$.
Such an approach, without resort to the CRAB smoothness variational procedure but with a possibly larger set of CD terms, is known in the literature as Digitized counter-diabatic QAOA (DC-QAOA)~\cite{ChandaranaDigitized2022}. 
DC-QAOA has been extensively used in a wide range of systems, including Ising models and classical optimization problems~\cite{ChandaranaDigitized2022}, MAXCUT and Sherrington-Kirkpatrick model~\cite{LiuEfficient2025}, Protein Folding~\cite{ChandaranaDigitized2024}, Molecular docking~\cite{DingMolecular2024}, and Bin packing model~\cite{XuDigitized2025}.
\end{description}

\section{Results} \label{sec:Results}
The performance of the different approaches is quantified and compared primarily using the rescaled residual energy~\cite{RuiyiExponential2025}:
\begin{equation} \label{eqn:epsilon_res}
\epsilon^{\res}_{\Pscript} = 
\frac{E_{\Pscript}(\btheta^{\opt}) -E_{\mathrm{gs}} }{\Nsites} \;,
\end{equation}
where $E_{\Pscript}$ is the expectation value of the target Hamiltonian $\Ho_z$ computed using the optimal solutions $\ket{\psi_{\Pscript}(\btheta^{\opt})}$ obtained with the different methods 
\begin{equation}
E_{\Pscript}(\btheta^{\opt}) =\bra{\psi_{\Pscript}(\btheta^{\opt})} \Ho_{z}\ket{\psi_{\Pscript}(\btheta^{\opt})} \;,
\end{equation}
and $E_{\mathrm{gs}}=-(\Nsites-3)J-2J_w+J_f$ denotes the ground state energy of $\Ho_z$.
By definition, $\epsilon^{\res}_{\Pscript}\ge 0$: it vanishes only if the optimal solution is the ground state of $\Ho_{z}$. 
For all the results, we work in units where $J=1$ and $\hbar=1$.

\begin{figure}[H]
    \centering
    \includegraphics[width=0.98\columnwidth]{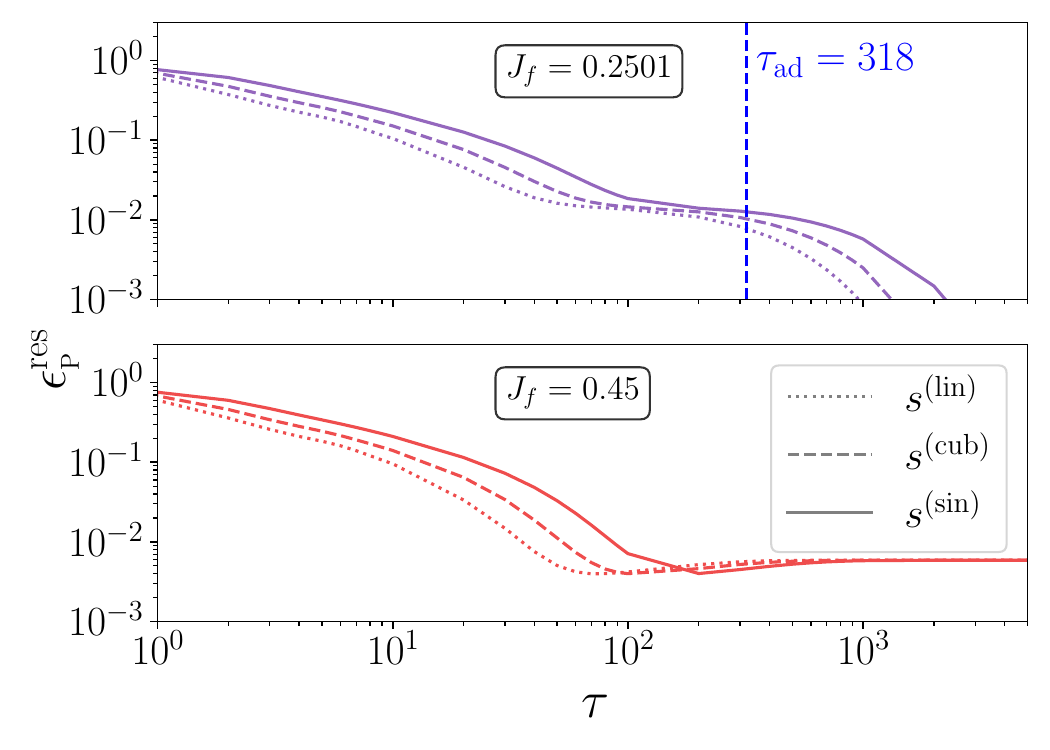}
    \caption{Residual energy calculated following nearly continuous-time evolution, by setting $\Deltat=0.1$ in Eq.~\eqref{eqn:U_dQA}, under fixed schedules --- namely, the linear ramp $s^{(\mathrm{lin})}(t)$, cubic ramp $s^{(\mathrm{cub})}(t)$, and sinusoidal ramp $s^{(\mathrm{sin})}(t)$ --- as defined in Eqs.~(\ref{linearrampeqn}), (\ref{cubicrampeqn}), and (\ref{sinerampeqn}), respectively. 
    We considered a ring with size $\Nsites=17$ and a weak (top, $J_f = 0.2501$) or strong (bottom, $J_f = 0.45$) frustration. $\tau_{\mathrm{ad}}\approx 318$ (top figure) corresponds to the minimum energy gap obtained for $J_{f}=0.2501$. For $J_{f}=0.45$, one predicts a value of $\tau_{\mathrm{ad}}\approx 0.96\times 10^{13}$.}
\label{ResidualEnergy_with_fixed_schedule_N_17}
\end{figure}

\subsection{Difficulty of ordinary fixed-schedule QA}
To appreciate the difficulty associated with the spin-glass-like bottleneck present in our frustrated Ising ring model, let us start with the most basic QA strategy \textbf{0)}, based on a fixed schedule $s(t)$ without CD terms. 

Figure~\ref{ResidualEnergy_with_fixed_schedule_N_17} shows the residual energy obtained from fixed-schedule annealing of duration $\tau$, using the evolution operator in Eq.~\eqref{eqn:U_dQA} with the linear ramp in Eq.~\eqref{linearrampeqn}, the cubic ramp in Eq.~\eqref{cubicrampeqn}, and the sinusoidal ramp in Eq.~\eqref{sinerampeqn}. 
Here, we consider a frustrated Ising ring of size $\Nsites=17$, with nearly continuous-time evolution generated by setting $\Deltat=0.1$. 
The upper panel shows the weakly frustrated case with $J_f=0.2501$, while the lower one corresponds to the highly frustrated case with $J_f=0.45$. 
Regardless of the value of $J_f$, the performance of all three schedules is comparable, with only marginal differences. 
In the case of weak frustration, the residual energy exhibits an exponentially decreasing behavior only after a time $\tau_{\mathrm{ad}}\approx 318$, which corresponds to the inverse square of the minimum energy gap shown in Fig.~\ref{MinEneVsN} for $\Nsites=17$: curiously, the vanishing derivatives of the cubic and sinusoidal schedule do not appear to lead to a lower residual energies in these regimes of $\tau$. 
For $J_f=0.45$, the residual energies of the different schedules have a nonmonotic increase toward a very long plateau, which is expected to decline exponentially only after $\tau_{\mathrm{ad}}\approx 0.96\times 10^{13}$, a timescale far beyond the range of Fig.~\ref{ResidualEnergy_with_fixed_schedule_N_17}. 
In conclusion, for the model considered, strategy \textbf{0)}, i.e., QA with a fixed schedule, appears to be quite inefficient, especially in the highly frustrated case. 

\begin{figure}[htb]
\centering
\includegraphics[width=\columnwidth]{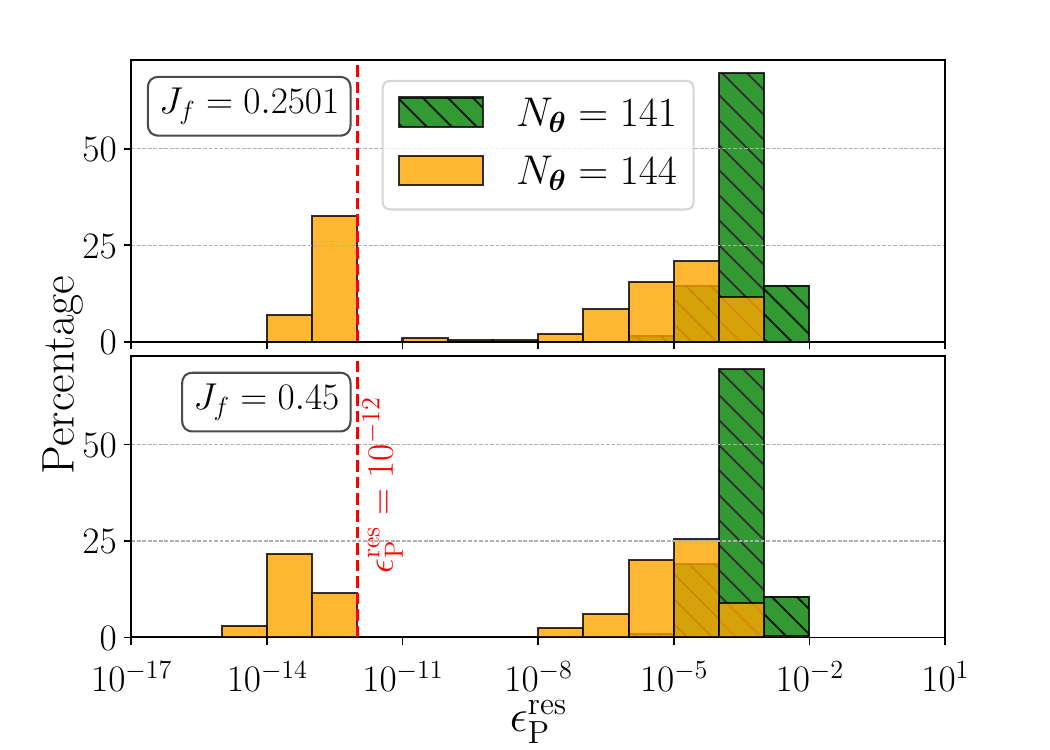}
\caption{
The histogram of the residual energy for a fixed system size $\Nsites = 17$ and for different frustrated couplings $J_f$. A total of $200$ randomly initialized sets of variational parameters $\boldsymbol{\theta}$ are optimized for a fixed number of layers $\Ptrot = 47$ and $48$, corresponding to a total number of parameters $N_{\btheta} = 3 \Ptrot = 141$ and $144$, respectively. The percentage of optimization runs (out of 200) yielding residual energies within a given range is indicated by the vertical bars. The red dashed vertical line indicates the prescribed threshold to identify the optimization runs that lead to the exact ground state.}
\label{DC_QAOA_histogram}
\end{figure}

\subsection{Controllability of the system using DC-QAOA}
Next we consider the issue of the {\em controllability} of the frustrated ring model~\cite{ArezzoDigital2025}, which, within the DC-QAOA {\em Ansatz}, is predicted to satisfy Eq.~\eqref{eqn:control_3P}.

Figure~\ref{DC_QAOA_histogram} shows histograms of the residual energy $\epsilon^{\res}_{\Pscript}$, see Eq.~\eqref{eqn:epsilon_res}, where $E_{\Pscript}$ is computed by optimizing the {\em Ansatz} state in  Eqs.~\eqref{eqn:U_CD}-\eqref{eqn:psi_P_CD}.
To construct the histograms, we fix the system size to $\Nsites=17$ --- corresponding to a dimensionality of gaussian states given by $\dim_{\mathrm{Gaussian}}^{\text{symm}}  =
(\Nsites^2 -1)/2=144$ ---, set the frustrated coupling to either $J_f=0.2501$ or $J_f=0.45$, and initialize the optimization of the $N_{\btheta} = 3\Ptrot$ parameters $\btheta$ from random values, uniformly sampled in the interval $[0,2\pi]$, for two values of $\Ptrot$, $\Ptrot=47$ and $48$, corresponding to a total number of parameters $N_{\btheta} = 3\Ptrot = 141$ and $144$, respectively.  
We perform $200$ independent optimization runs and record the resulting residual energy from each run. The bars of the histogram represent the percentage of optimization runs (out of $200$) that yield residual energies within a specified range.
Taking the numerical precision of the optimization routines into account, we set a threshold of $10^{-12}$ and regard residual energies below this value as corresponding to the exact ground state of the frustrated Ising ring.~\cite{RuiyiExponential2025,ArezzoDigital2025}.

These histograms numerically validate the analytical expression, see Eq.~\eqref{eqn:control_3P}, for the critical number of parameters $N^{\mathrm{cr}}_{\btheta}=(\Nsites^2-1)/2$ required to achieve full controllability of the system, here $N^{\mathrm{cr}}_{\btheta}=144$, at which point a finite percentage of the optimization runs converge to the exact ground state, as indicated by residual energies below the prescribed threshold. 
This explicitly demonstrates the quadratic-in-$\Nsites$ controllability~\cite{ArezzoDigital2025} of the frustrated Ising ring for DC-QAOA protocols incorporating the lowest-order AGP. 
This full controllability is determined solely by the free fermionic nature, the underlying symmetries, and the size of the system, and it is independent of the spectral gap of the interpolating Hamiltonian and, consequently, of the annealing difficulty. 
This is evident from the identical critical value of $N^{\mathrm{cr}}_{\btheta}$ obtained for both values of $J_f$, ranging from a weakly frustrated case at $J_f = 0.2501$ to a strongly frustrated value $J_f = 0.45$. 

Given that the critical number of variational parameters $\btheta$ is independent of the exponentially small instantaneous gap induced by the frustration, and considering the increasing difficulty of QA with stronger frustration, 
in the following, we will primarily focus on systems with $J_f = 0.45$.

\begin{figure}[htb]
\centering
\includegraphics[width=\columnwidth]{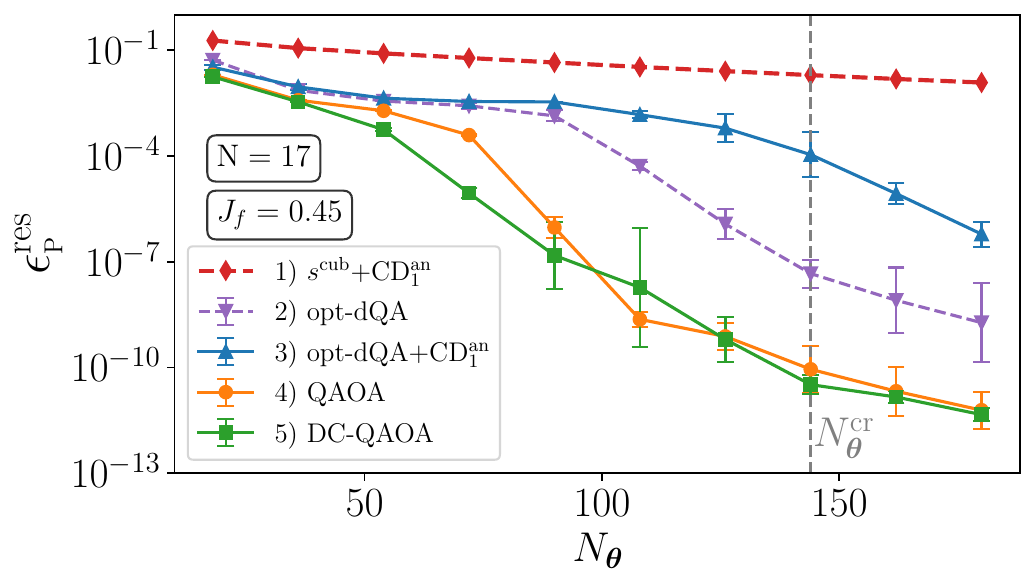}\\
\includegraphics[width=\columnwidth]{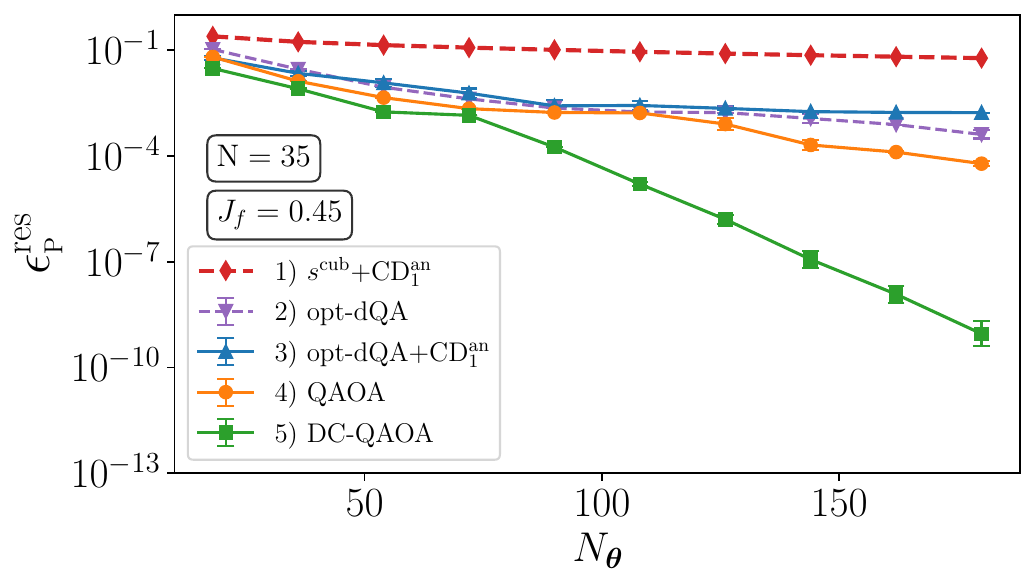}
\caption{The residual energies obtained from different optimization techniques for $\Nsites=17$ (top) and $\Nsites=35$ (bottom) with circuits of different depth $N_{\btheta}$.
For the methods involving dCRAB, we fix the value of $\Nbasis = \Ptrot$. 
Furthermore, for each iteration of dCRAB, the optimization of the parameters $\C$ is performed $10$ times using randomly generated frequencies $\omega^{(1,2)}_{n}$. 
The set of optimized parameters corresponding to the lowest residual energy is then selected and treated as a single numerical experiment. The residual energy of dCRAB-incorporated methods shown in the figure is the average of best residual energies obtained from such $10$ numerical experiments and error bars correspond to $\pm 1$ standard deviation in log space. For $\Nsites=17$ the critical number of gates $N^{\mathrm{cr}}_{\btheta}=144$ is indicated using a gray dashed vertical line, while the corresponding critical number of gates for $\Nsites=35$ is $N^{\mathrm{cr}}_{\btheta}=612$, hence beyond the explored range of $N_{\btheta}$.
}
\label{fig:eps_res_methods}
\end{figure}

\subsection{Comparing five variational {\em Ansatzes}} \label{sec:comparison}
Next, we present the results obtained with the five different digital optimization strategies, both with and without local counter-diabatic driving, discussed in Sec.~\ref{sec:methods}, which we recap here:
\textbf{1)} A fixed cubic schedule $s_p$ with analytical CD terms; \textbf{2)} a CRAB-optimized schedule $s_p$ without, or \textbf{3)} with analytical CD terms; \textbf{4)} a QAOA {\em Ansatz} with CRAB-optimized parameters; \textbf{5)} a DC-QAOA {\em Ansatz} with CRAB-optimized parameters, including variational CD terms.  
In all relevant cases, the numerical optimization of the parameters  
is carried out using the Broyden–Fletcher–Goldfarb–Shanno (BFGS)~\cite{NocedalNumerical2006} algorithm, together with the corresponding analytical gradient of the expectation value of the energy. 
This approach ensures efficient convergence towards the global minimum or a near-global local minimum.

Figure ~\ref{fig:eps_res_methods} shows the rescaled residual energies $\epsilon_{\Ptrot}^{\res}$ obtained by the various methods for $\Nsites=17$ (top) and $\Nsites=35$ (bottom) as a function of the {\em circuit depth} $N_{\btheta}$: more precisely, $N_{\btheta}=2\Ptrot$ for QAOA --- method \textbf{4)} --- and $N_{\btheta}=3\Ptrot$ for DC-QAOA --- method \textbf{5)}; moreover, conventionally, we set $N_{\btheta}=2\Ptrot$ for all methods without CD driving --- $\theta^{xz}_p=0$, hence also for method \textbf{2)} --- while $N_{\btheta}=3\Ptrot$ for all methods with CD driving --- hence also for method \textbf{1)} and \textbf{3)}. For the digital fixed-schedule dynamics in methods \textbf{1)-3)}, we set $\Deltat=1$, since optimizing $\Deltat$ consistently yields values close to unity.
For a single numerical experiment of methods \textbf{2)-5)} involving CRAB-optimization with random frequencies, we repeat the optimization of the Fourier coefficient parameters $\C$ starting from 10 samples of different random frequencies, and select the optimized parameters corresponding to the best residual energy. Moreover, to estimate the error bars, we evaluate $\epsilon_{\Ptrot}^{\res}$ for 10 times, and report the geometric average of the various $\epsilon_{\Ptrot}^{\res}$ with error bars corresponding to $\pm 1$ standard deviation in log space.

In general, the relative performance of the various techniques is quite independent of the system size. The fixed schedule method \textbf{1)} exhibits the poorest performance, totally unable to negotiate the difficult spectral gap bottleneck. 
A CRAB-optimization of the schedule $s_p$, method \textbf{2)}, improves the performance, but the addition of analytical CD terms, method \textbf{3)}, appears to be counter-productive. 
QAOA-based approaches, method \textbf{4)} and \textbf{5)} have similar performance for small sizes ($\Nsites=17$), with a clear performance improvement brought in by the addition of optimized variational CD terms, which is quite clear for larger sizes ($\Nsites=35$). 
The effect of different levels of frustration on the residual energy of the two most efficient methods, QAOA and DC-QAOA, is shown in Appendix~\ref{app:EreswithdiffJf}. 

\begin{figure}[htb]
\centering
\includegraphics[width=\columnwidth]{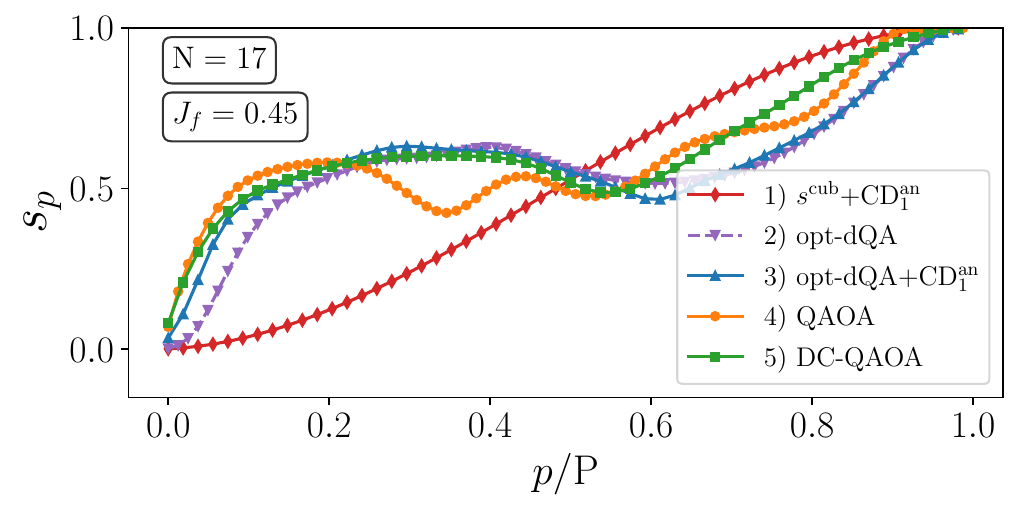}
\includegraphics[width=\columnwidth]{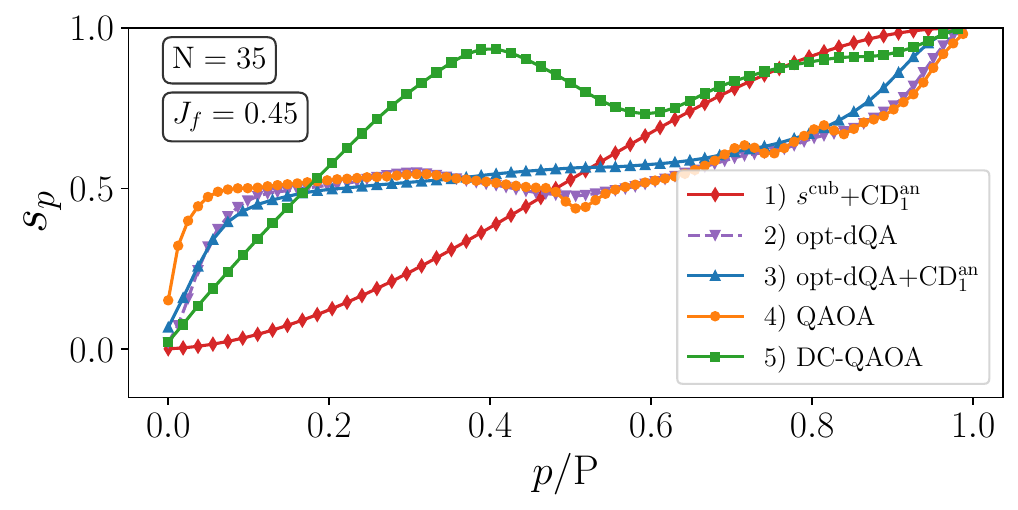}
\caption{The schedule parameter $s_p=\theta^z_p/(\theta^x_p+\theta^z_p)$ of the various methods, for $\Nsites=17$ (top) and $\Nsites=35$ (bottom), with a fixed total number of parameters $N_{\btheta}=162$.
}
\label{fig:sp_methods}
\end{figure}

\begin{figure}[htb]
\centering
\includegraphics[width=\columnwidth]
{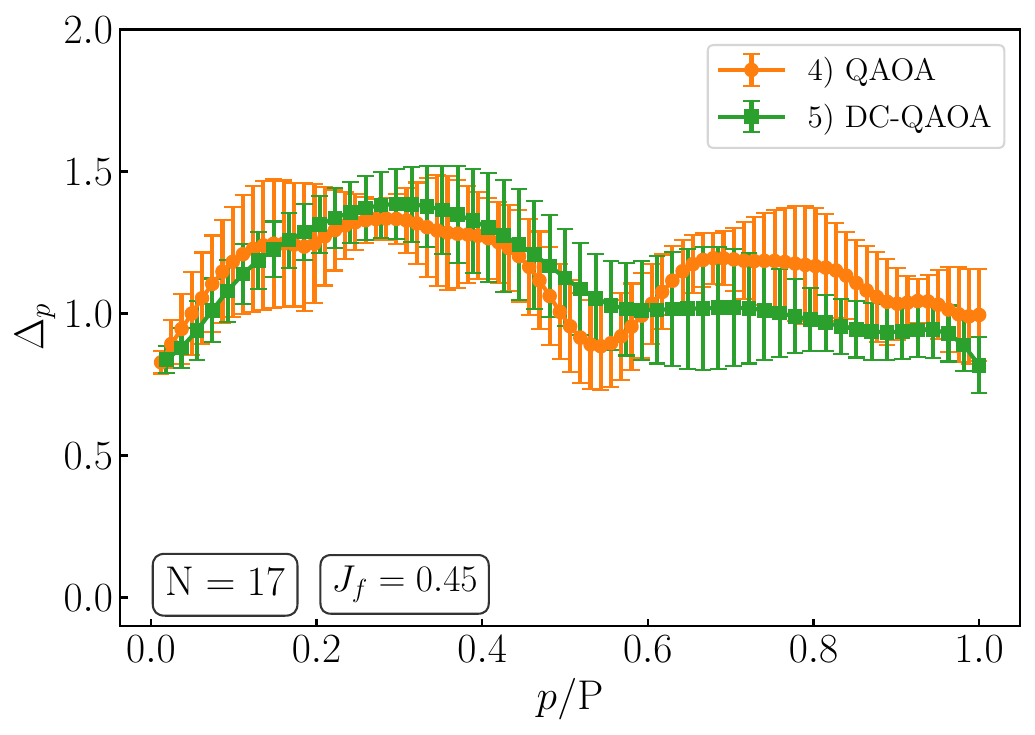}
\includegraphics[width=\columnwidth]
{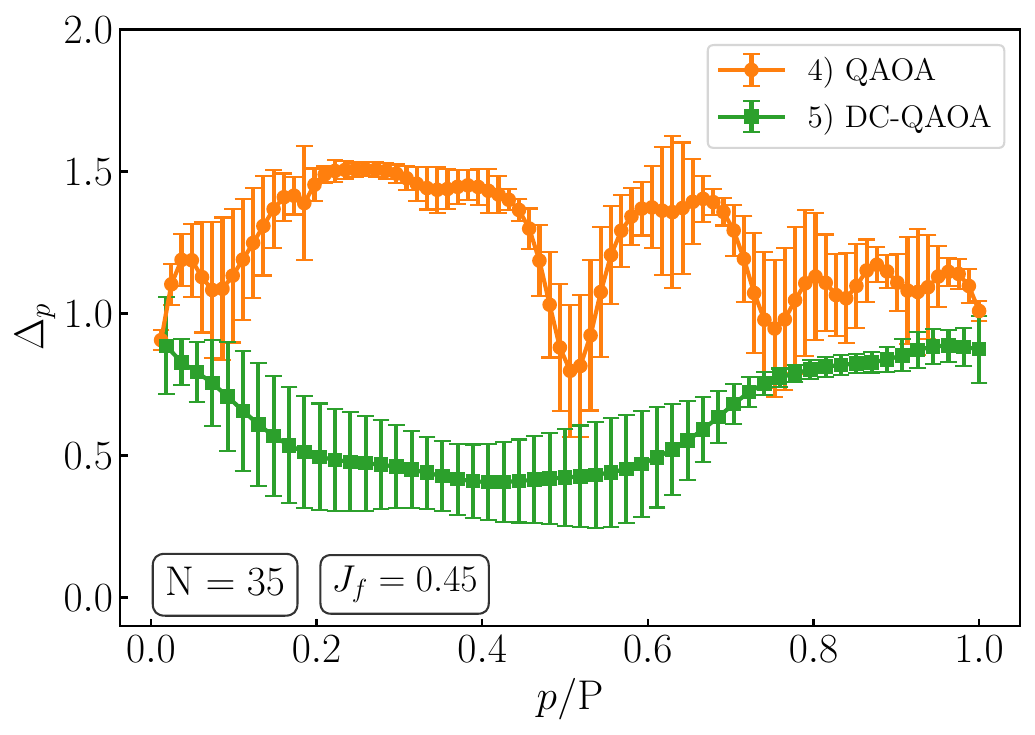}
\caption{The optimal ``time-steps'' $\Delta_p=\theta^x_p+\theta^z_p$ of QAOA and DC-QAOA, for $\Nsites=17$ (top) and $\Nsites=35$ (bottom). Here $N_{\btheta}=162$.
}
\label{fig:TimeSteps_methods}
\end{figure}

\begin{figure}[htb]
\centering
\includegraphics[width=\columnwidth]
{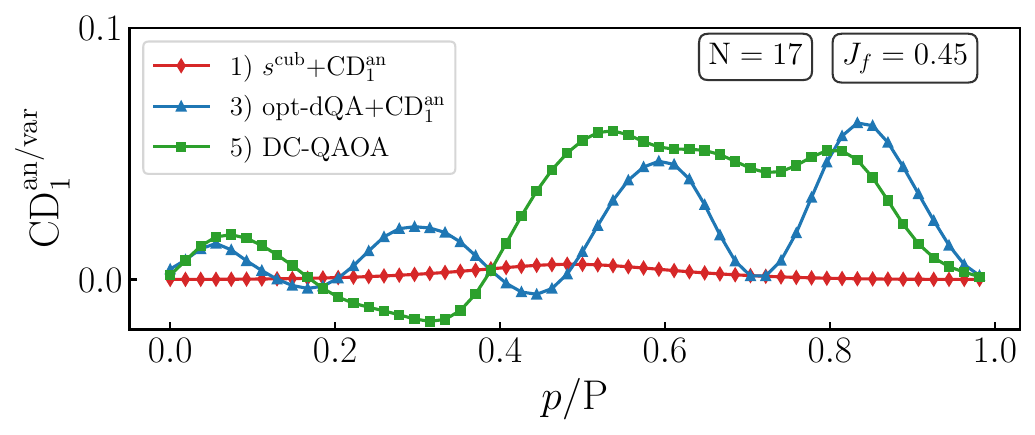}
\includegraphics[width=\columnwidth]{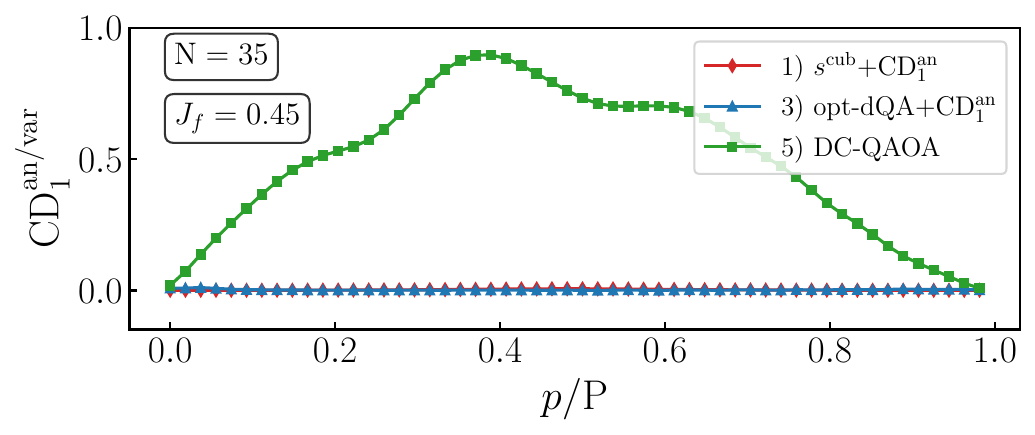}
\caption{The counterdiabatic parameter $\CDan$ or $\CDvar$ of the various methods, see Eq.~\eqref{eqn:CD_params}, for $\Nsites=17$ (top) and $\Nsites=35$ (bottom), with a fixed total number of parameters $N_{\btheta}=162$.
}
\label{fig:CD_methods}
\end{figure}

\begin{figure}[htb]
\centering
\includegraphics[width=\columnwidth]
{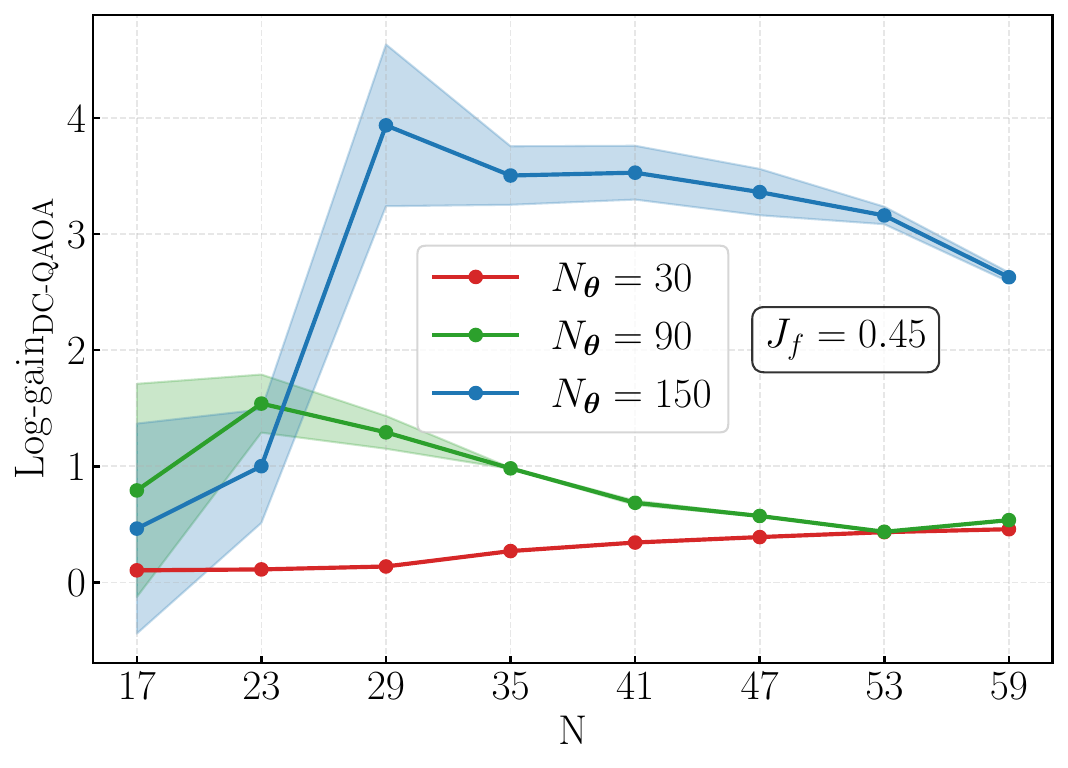}\label{Log_Improvement_Mean_Std_DC_QAOA_vs_QAOA_MultipleNG_Jf_0.45}
\caption{The log-gain of residual energies, defined in Eq.~\eqref{eqn:Log-gain}, which compares the performance of the QAOA and DC-QAOA methods, versus the system size $\Nsites$, with three different fixed circuit depths $N_{\btheta}$. The CRAB specifications and the procedure behind the numerical experiments remain the same as those in Fig.~\ref{fig:eps_res_methods}.}
\label{fig:LogGain}
\end{figure}

Figures \ref{fig:sp_methods}, \ref{fig:TimeSteps_methods} and \ref{fig:CD_methods} give some insight on the variational parameters $\btheta$ behind the various methods, showing the effective schedule $s_p$ and time-step $\Delta_p$
\begin{align} \label{eqn:sp_params}
    s_p = \frac{\theta^z_p}{\Delta_p} \;, \hspace{15mm} 
    \Delta_p = \theta^x_p+\theta^z_p \;,
\end{align}
and the analytical or variational counter-diabatic parameters related to $\theta^{xz}_p$:
\begin{align} \label{eqn:CD_params}
\CDan(s_p) &= \frac{\theta^{xz}_p}{\Deltat} = \dot{s}_{p}\alpha^{\opt}_{1}(s_{p}), \nonumber \vspace{3mm} \\
\CDvar &= \frac{\theta^{xz}_p}{\Deltap} \;.
\end{align}
Figure \ref{fig:sp_methods} shows that $s_p$ is strongly modified from the fixed cubic schedule used in method \textbf{1)}. 
A finer comparison of the two QAOA-methods \textbf{4)-5)} shows a noticeable difference in $s_p$ and $\Delta_p$ for larger sizes ($\Nsites=35)$. Concomitantly, see Fig.~\ref{fig:CD_methods}, the variational CD-related parameter 
$\CDvar = \theta^{xz}_p/\Deltap$, which is quite small for $\Nsites=17$, becomes appreciable and evidently important in the good performance of DC-QAOA for $\Nsites=35$. 

We now analyze in more detail the best two techniques: \textbf{4)} QAOA and \textbf{5)} DC-QAOA. 
The dependence of the circuit expressivity on the number of parameters and the system size is illustrated in Fig.~\ref{fig:LogGain} through the logarithmic gain:
\begin{equation} \label{eqn:Log-gain}
\text{Log-gain}_{\text{DC-QAOA}} = \mathrm{log}_{10}\left(\frac{\epsilon^{\res}_{\Pscript}~\text{of QAOA}}{\epsilon^{\res}_{\Pscript}~\text{of DC-QAOA}}\right) \;,
\end{equation}
which quantifies the improvement in residual energy in terms of orders of magnitude achieved by DC-QAOA relative to QAOA. 
With small number of parameters ($N_{\btheta}=30$) the circuit is less expressive, still it shows a nominal gain with counter-diabaticity, especially for the large system sizes. 
The superiority of DC-QAOA persists with increasing system size and number of parameters $N_{\btheta}$. 
However, when $N_{\btheta}$ approaches or exceeds its critical value, the performance of the two methods becomes comparable. 
This is reflected in Fig.~\ref{fig:eps_res_methods}(top) and in the reduced values of the Log-gain for $\Nsites = 17$ and $\Nsites = 23$ at $N_{\btheta}=90$ and $N_{\btheta}=150$. 
In this neighborhood of $N^{\mathrm{cr}}_{\btheta}$, the larger standard deviation of the Log-gain indicates that the inclusion of counter-diabaticity has only a marginal effect, and any resulting performance improvement is inconsistent.

\begin{figure}[htb]
\centering
\includegraphics[width=\columnwidth]
{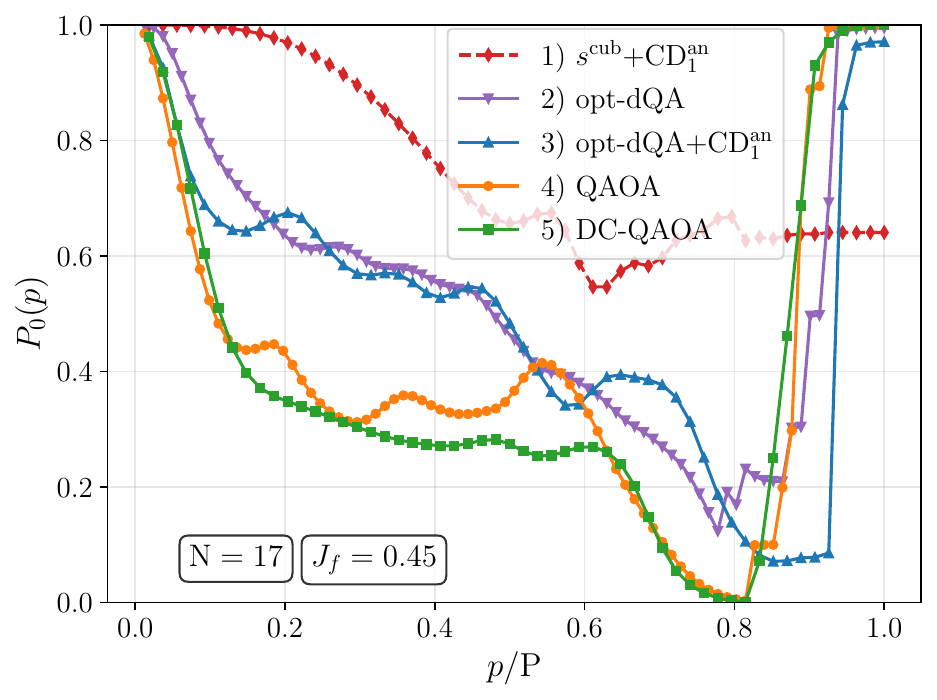}
\includegraphics[width=\columnwidth]
{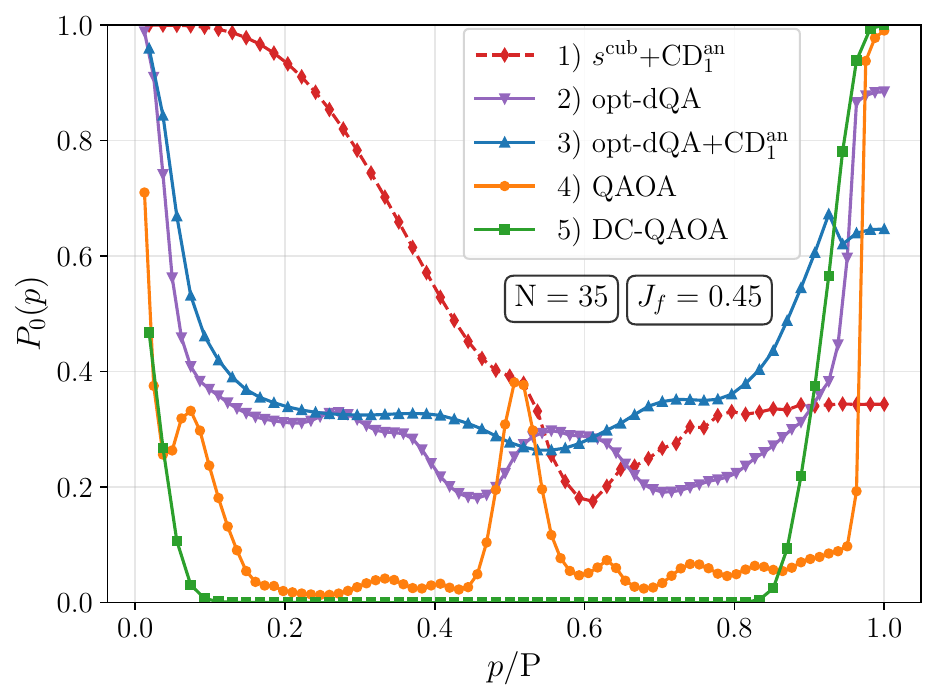}
\caption{The instantaneous ground state fidelity of different strategies with smooth schedules for systems of size $\Nsites=17$ (top) and $\Nsites=35$ (bottom). The methods illustrated and the corresponding specifications of the numerical experiments are same as that of Fig.~\ref{fig:eps_res_methods} for a fixed total number of parameters $N_{\btheta}=162$. The ground-state fidelity shown is the average of all $10$ trials of the numerical experiments.
}
\label{fig:Population_methods}
\end{figure} 
 
\subsection{Populations on the instantaneous ground state} \label{sec:populations}
The dynamics of the various optimization strategies throughout the evolution process is revealed by the instantaneous eigenstates population~\cite{RuiyiExponential2025}:
\begin{equation}
    P_{j} (p) = \vert\langle\phi^{j,\mathrm{eff}}_{p}\vert\psi_{p}\rangle\vert^{2},
\end{equation}
where 
\begin{equation}
\ket{\psi_{p}}=\prod_{m=1}^{p}\e^{-\frac{i}{\hbar}\theta^{xz}_{m}\Ho_{xz}}\e^{-\frac{i}{\hbar}\theta^x_{m}\Ho_x}\e^{-\frac{i}{\hbar}\theta^z_{m}\Ho_z}\ket{\psi_0}
\end{equation}
is the exact evolved state of the system at each step $p$, and $\ket{\phi^{j,\mathrm{eff}}_{p}}$ is the $j^{\text{th}}$ instantaneous eigenstate of the effective Hamiltonian~\cite{RuiyiExponential2025}:  
\begin{equation}
    \Ho^{\mathrm{eff}}_p = s_p\Ho_z + (1-s_p)\Ho_x \;.
\end{equation}
The population of the eigenstates, particularly that of the ground state, indicates whether the discrete finite-time dynamics of the various strategies remain close to the ideal annealing (adiabatic) dynamics or whether the optimization realizes shortcuts to the target state through the excited states.

Figure \ref{fig:Population_methods} shows the instantaneous population of the ground state $P_0(p)$, for different optimization strategies --- whose residual energies are discussed in Fig.~\ref{fig:eps_res_methods} ---, for systems with size $\Nsites=17$ (top) and $\Nsites=35$ (bottom). 
In general, all the methods show excitations away from the ground state, setting them far from the ideal adiabatic dynamics of the effective Hamiltonian. However, strategies optimized with CRAB regain the ground state probability towards the end of the evolution. Here, the general tendency is to find a better path to the target state through the excited states, which is expected for finite-time dynamics. 

Notice that, for $\Nsites = 35$, the considerable involvement of $\Ho_{xz}$ leads to quick excitation to higher energy states, leaving a negligible chance to be in the ground state. 
Therefore, in this context, the additional first-order local counter-diabatic term explores better shortcuts to the target state using effective intermediate transitions.
This is, in some sense, a curious manifestation of the {\em law of unintended consequences}: in a continuous-time framework, the counter-diabatic terms should, in principle, keep the system in the ground state, while here, when used as additional variational parameters in a digitized framework, they indeed improve the performance of DC-QAOA by further depressing the ground state population at the early stages of the digital dynamics, hence performing an effective shortcut-to-adiabaticity. 

\section{Discussion and Conclusions}
\label{sec:conclusions}
We have explored different discrete-time optimization techniques with and without local counter-diabatic (CD) terms, and applied them to a frustrated Ising ring model, which exhibits an exponentially small energy gap --- a bottleneck that challenges a conventional Quantum Annealing approach. 

We find that fixed schedule techniques perform very poorly, and the addition of local lowest-order CD terms does not lead to any further improvement. Schedule optimization, by a Fourier-based CRAB improves the results: once again, the addition of local lowest-order CD terms is not useful.
A clear step forward in the performance is obtained by QAOA with CRAB-optimized smooth parameters. 
A further improvement is obtained by including a QAOA layer obtained by lowest-order CD terms but with {\em free variational parameters}, once again CRAB-optimized: this method, often named DC-QAOA\cite{ChandaranaDigitized2022}, is the best performing one, on an equal layer-depth basis, particularly for large system sizes.  

The controllability of the frustrated Ising ring in the presence of the local CD terms is examined through the bare optimization of the DC-QAOA parameters. 
We verified that the system becomes fully controllable with a total number of gates that scales quadratically with the system size. 
Therefore, the analytical results of Ref.~\cite{ArezzoDigital2025} in the context of QAOA are also validated for DC-QAOA, since the free-fermionic nature and the symmetries of the unitary operators generated by $\Ho_z$ and $\Ho_x$ are preserved by those of the first-order AGP with operator part $\Ho_{xz}$.

A quite important observation is that the variationally optimized CD terms effectively perform a {\em shortcut-to-adiabaticity} towards the final target state: this is strange, at a first sight, because the CD mechanism is in principle intended to maintain the evolved state on the instantaneous ground state, rather than effectively provoking transitions to higher states. 
This simply tells us that the CD terms at the level of a digitized evolution are really not doing what one would expect for a continuous-time evolution~\cite{BerryTransitionless2009,SelsMinimizing2017,KOLODRUBETZ20171}.

One thing we have not discussed, which would be worth addressing in future studies, is the role of local bias longitudinal fields, as introduced in Ref.~\cite{CadavidBias2025}: in combination with a fixed sinusoidal schedule and local CD-terms, they are at the basis of a method called DCQO~\cite{HegadeDigitized2022}. 
Local bias longitudinal fields effectively rotate the starting initial state $|\psi_0\rangle=|+\rangle^{\otimes \Nsites}$ into the $x-z$ plane in spin space, and have been shown to considerably improve the convergence of QAOA~\cite{YuQuantum2022}. 
In the present paper, we have not explicitly included local bias fields for two reasons. The first reason is technical: we heavily exploited the Jordan-Wigner mapping to spinless fermions, in which local longitudinal fields cannot be addressed because of a non-local Jordan-Wigner string term which would appear in the fermionic Hamiltonian. The second reason is more physical, the ground state of our frustrated Ising ring model is a rather dull ferromagnet, with all spins aligned: obviously, longitudinal bias field would rather quickly point towards the ferromagnetic ground state.
More challenging, and hence left for future studies, is the inclusion of local longitudinal fields in DC-QAOA, applied to non-trivial computational problems like MaxCut~\cite{ZhouQuantum2020,Pecci_2024}, where the spin-glass ground state is hard to find. 

\begin{acknowledgments}
G.E.S. acknowledges financial support from PNRR MUR project PE0000023-NQSTI, from PRIN 2022H77XB7 of the Italian Ministry of University and Research, and from the QuantERA II Programme STAQS project that has received funding from the European Union’s H2020 research and innovation programme under Grant Agreement No 101017733.

\end{acknowledgments}
\section*{Data Availability}
The data that support the findings of this study are archived on Zenodo (see Ref.~\cite{CD-QAOAData2026}) and mirrored on GitHub.

\appendix
\begin{widetext}

\section{Technical details on the various methods}
\label{app:details}
\subsection{Analytical CD terms}
We analytically optimize the parameter $\alpha_{1}$ of Eq.~\eqref{AGP1st} by minimizing the action~\cite{HegadeDigitized2022,WurtzCounterdiabaticity2022}
\begin{equation}
    \calS(\calA^{(1)}_{s})= \Trace\big[G^2_{s}\big],
\end{equation}
where $G_{s}=\partial_{s}\Ho(s)+\frac{i}{\hbar}\left[\calA^{(1)}_{s},\Ho(s)\right]$ and $\Ho(s)$ is given by Eq.~\eqref{InterpHamil}. The minimization of $\calS$ results in an optimal value of $\alpha_1$:
\begin{align}
\alpha^{\opt}_{1}(s_p) 
= \frac{\hbar \, \Trace ([\Ho(s_p),\partial_s\Ho])^2}{\Trace([\Ho(s_p),[\Ho(s_p),\partial_s\Ho]])^2}
= -\frac{\sum_j J_j^2}{16 (1-s_p)^2 \sum_j J_j^2 + s_p^2( 12 \sum_j J_j^2 J_{j+1}^2 + 4 \sum_j J_j^4 )} \;,
\end{align}
where the last expression applies to our Ising model. 

\subsection{The dressed-CRAB schedule}
For the optimization of the schedule in Eq.~\eqref{eqn:s_p_CRAB}, we adopt the dressed-CRAB (dCRAB)~\cite{RachDressing2015} technique with two super-iterations.
In the first dCRAB iteration, the schedule is initialized with Fourier frequencies $\omega^{(1)}_{n}$:
\begin{equation}
    s^{(1)}_p  = \frac{t_p}{\tau} +\frac{t_p}{\tau} \sum_{n=1}^{\Nbasis} \rmC_n \sin(\omega^{(1)}_n (t_p-\tau))\;.
    \label{s1p}
\end{equation}
We optimize the parameters $\C=(\rmC_1,....,\rmC_\Nbasis)$ by minimizing
\begin{equation}
    \begin{split}
        E_{\Pscript}(\btheta(\C)) = \bra{\psi_{\Pscript}(\btheta(\C))} \Ho_{z}\ket{\psi_{\Pscript}(\btheta(\C))} \;,
    \end{split}
    \label{ExpctEnelocalCD}
\end{equation}
where $\btheta(\C)$ is given by Eq.~\eqref{eqn:theta_Opt_sp}.
At the beginning of the optimization, we set $\rmC_{n}=0$ to initialize $s^{(1)}_p$ as $t_p/\tau$.
Next, we improve $s^{(1)}_p$ with a second dCRAB iteration by adding another $\Nbasis$ Fourier modes to it:
\begin{equation}
    s^{(2)}_p  = s^{(1)}_p +\frac{t_p}{\tau} \sum_{n=1}^{\Nbasis} \rmC^{(2)}_n \sin(\omega^{(2)}_n (t_p-\tau))\;,
    \label{s2p}
\end{equation}
and optimizing the new set of parameters $\C=(\rmC^{(2)}_1,....,\rmC^{(2)}_\Nbasis)$.
Following Ref.~\cite{RuiyiExponential2025}, we keep the value of $\Nbasis = \Ptrot$ and choose the Fourier frequencies $\omega^{(1,2)}_{n}$ by generating $x=\omega^{(1,2)}_n\tau/\pi$ drawn from the Gamma probability distribution:
\begin{align}
    \Gamma(x\ge 0; \alpha, \beta) = \dfrac{x^{\alpha-1}\e^{-x/\beta}}{\beta^{\alpha}\Gamma(\alpha)} \;,
    \label{betadistr}
\end{align}
where $\alpha>0$ and $\beta>0$ are shape parameters and $\Gamma(\alpha)$ denotes the Euler Gamma function. We took $\alpha = 3/2$ and $\beta = 4$ to generate $\omega^{(1)}_{n}$, and $\alpha = 3/2$ and $\beta = 20$ to generate $\omega^{(2)}_{n}$.

\textbf{Adding analytical CD:} 
The CD terms corresponding to the dCRAB schedules are included in the optimization routine by calculating $\alpha^{\opt}_{1}(s_{p})$ and $\dot{s}_p$ for the schedules in Eqs.~(\ref{s1p}) and (\ref{s2p}). 
Conceptually, this approach is closely related to the counter-diabatic optimized local driving method~\cite{IevaCounterdiabatic2023}, where additional optimal control terms are used to realize efficient path to the target state. Our approach does not use such additional terms.

\subsection{QAOA with dressed CRAB}

We use dCRAB to optimize $\btheta$ given by Eq.~\eqref{eqn:theta_QAOA}, where in the first iteration, they are projected onto Fourier modes of frequencies $\omega^{(1)}_{n}$:
\begin{equation}
    \begin{split}
    \theta^{z,1}_p &= \rmC^{z}_0\frac{t_p}{\tau} +\frac{t_p}{\tau} 
    \sum_{n=1}^{\Nbasis} \rmC^{z}_n \sin(\omega^{(1)}_n (t_p-\tau)) 
    \vspace{2mm}, \\
    \theta^{x,1}_p
    &= \rmC^{x}_0\Big(1 - \frac{t_p}{\tau}\Big)
    +\Big(1 - \frac{t_p}{\tau}\Big) 
    {\sum_{n=1}^{\Nbasis}} \rmC^{x}_n \sin(\omega^{(1)}_n (t_p-\tau)),
    \end{split}
    \label{DcrabFirstIter}
\end{equation}
and optimize 
$\C=(\rmC_{0}^x,...,\rmC_{\Nbasis}^x,\rmC_{0}^z,...,\rmC_{\Nbasis}^z)$ through minimization of the expectation value given in Eq.~(\ref{ExpctEnelocalCD}).
At the beginning of the optimization, we set $\rmC^{z}_{0}=\rmC^{x}_0=1$ and $\rmC^{z}_{n}=\rmC^{x}_n=0$.

The obtained solution can be further improved by a second dCRAB iteration with another $\Nbasis$ Fourier modes added to $\theta^{z,1}_p$ and $\theta^{x,1}_p$:
\begin{equation}
    \begin{split}
    \theta^{z,2}_p &= \rmC^{z,2}_0\theta^{z,1}_p +\frac{t_p}{\tau} 
    \sum_{n=1}^{\Nbasis} \rmC^{z,2}_n \sin(\omega^{(2)}_n (t_p-\tau)) 
    \vspace{2mm}, \\
    \theta^{x,2}_p
    &= \rmC^{x,2}_0\theta^{x,1}_p
    +\Big(1 - \frac{t_p}{\tau}\Big) 
    {\sum_{n=1}^{\Nbasis}} \rmC^{x,2}_n \sin(\omega^{(2)}_n (t_p-\tau)),
    \end{split}
    \label{DcrabSecondIter}
\end{equation}
and optimize the new set parameters 
$\C=(\rmC_{0}^{x,2},...,\rmC_{\Nbasis}^{x,2},\rmC_{0}^{z,2},...,\rmC_{\Nbasis}^{z,2})$ using the same strategy of the first dCRAB iteration. We keep $\Nbasis=\Ptrot$, and select the Fourier frequencies from the Gamma function, as in the other dCRAB-incorporated methods in this work. 
The optimized $\btheta$ parameters are used to construct the smooth schedules~\cite{RuiyiExponential2025}:
\begin{equation}
    s_{p} = \frac{\theta^{z}_p}{\Deltap},
\end{equation}
with time steps
\begin{equation}
    \Deltap = \theta^{z}_p+\theta^{x}_p.
\end{equation}
In general, the above optimized time steps need not be uniform.

\textbf{Adding variational CD:} We optimize $\btheta$ given by Eq.~\eqref{eqn:theta_QAOA_CD} by projecting $\theta^{xz}_p$ onto a finite number $\Nbasis$ of smooth Fourier modes:  
\begin{equation}
    \begin{split}\theta^{xz,1}_p &=\rmC^{xz}_0\theta^{xz,0}_p+
    \sum_{n=1}^{\Nbasis} \rmC^{xz}_n \sin(\omega^{(1)}_n (t_p-\tau))
    \vspace{2mm}, \\
    \theta^{xz,2}_p &=\rmC^{xz,2}_0\theta^{xz,1}_p+
    \sum_{n=1}^{\Nbasis} \rmC^{xz,2}_n \sin(\omega^{(2)}_n (t_p-\tau)),
    \end{split}
    \label{txz1st2nd}
\end{equation}
along with the parameters defined in Eqs.~(\ref{DcrabFirstIter}) and (\ref{DcrabSecondIter}) during the first and the second dCRAB iterations, respectively. We consider no presence of $\theta^{xz,1}_p$ at the beginning of the optimization by setting $\theta^{xz,0}_p=\rmC^{xz}_{n}=0$.

\begin{figure*}[t]
    \centering
    \subfloat[]{\includegraphics[width=0.45\textwidth]{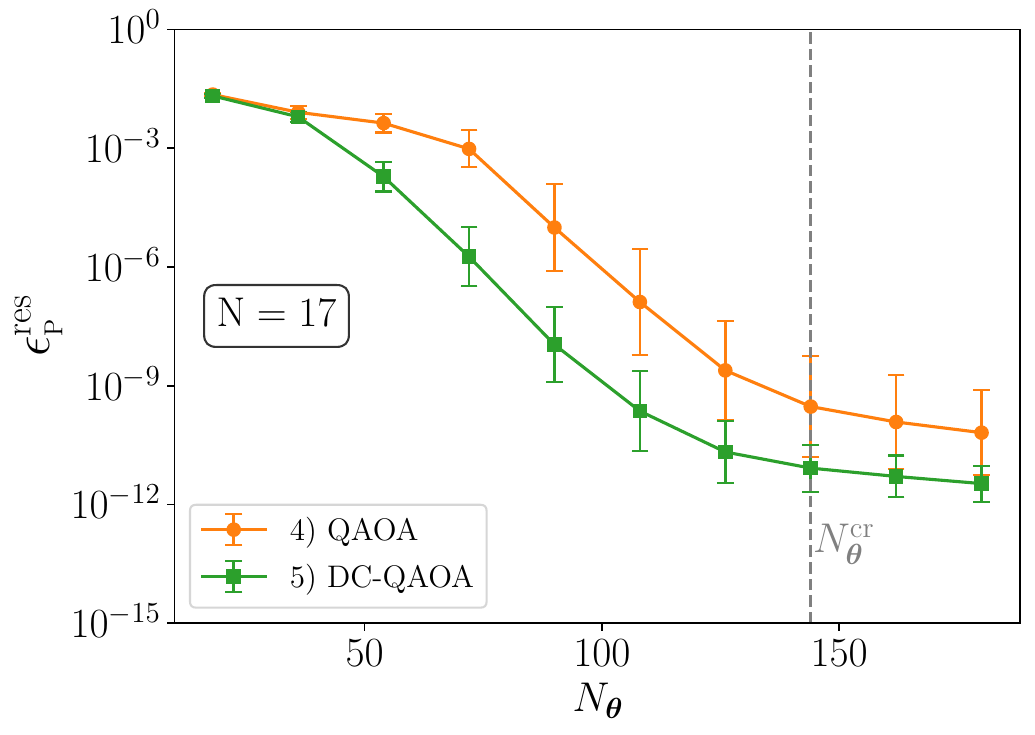}\label{Eres_Diff_Jf_N_17}}
    \hspace{0.02\textwidth}
    \subfloat[]{\includegraphics[width=0.45\textwidth]{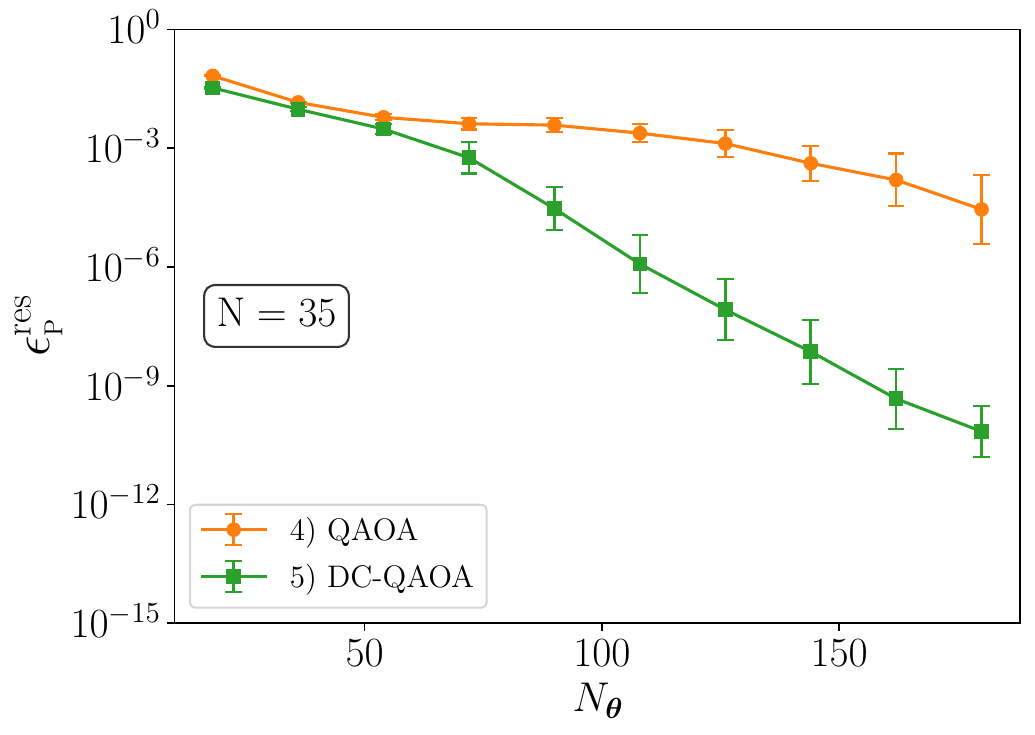}\label{Eres_Diff_Jf_N_35}}
    \caption{Residual energy obtained for a frustrated Ising ring of size (a) $\Nsites=17$ and (b) $\Nsites=35$ using 4) QAOA and 5) DC-QAOA, where the frustration is induced by randomly fixing $30$ different values of $J_f$ from the range $[0.2501,0.45]$. The dCRAB optimization routine and parameters are same as those of Fig.~\ref{fig:eps_res_methods}, except that only $1$ trial of the numerical experiment is executed for each value of $J_f$. The residual energy shown in the figure is the average of best residual energies obtained for all random values of $J_f$ and the error bars correspond to $\pm1$ standard deviation in log space.
    }
    \label{Eres_Diff_Jf}
\end{figure*}

\section{Residual energy for different frustrated couplings}
\label{app:EreswithdiffJf}
We can check the effect of the frustration on the two most effective methods; 4) QAOA and 5) DC-QAOA. Therefore, we randomly choose $30$ different $J_f$ values from a range $\left[0.2501,0.45\right]$ with uniform probability, which satisfy the condition $JJ_f>J^{2}_{w}.$ Further, we execute single numerical experiment of both methods for each value of $J_f$. For systems of sizes $\Nsites = 17$ and $\Nsites=35$, Fig.~\ref{Eres_Diff_Jf} shows the mean value of the obtained residual energies with different $J_f$ values, where the error bars indicate the standard deviation in log space. The superiority of the method 5) DC-QAOA is maintained across all levels of frustration and for both system sizes.

\end{widetext}

\bibliography{Biblio}
\end{document}